\def\preprint{0}               
\def\preprint{1}                
\DeclareRobustCommand{\ion}[2]{%
\relax\ifmmode
\ifx\testbx\f@series
{\mathbf{#1\,\mathsc{#2}}}\else
{\mathrm{#1\,\mathsc{#2}}}\fi
\else\textup{#1\,{\mdseries\textsc{#2}}}%
\fi}
\def\comment#1{}
\if\preprint1
        \documentclass[usenatbib]{mn2e}
        \usepackage{natbib}
        \usepackage{times}
        \usepackage{graphicx}
        \usepackage{calc}
\else
        \documentstyle[astrop-bib,referee,times]{mn}
        \newcommand{\includegraphics}[1]{}
\fi

\def\oversim#1#2{\lower0.5pt\vbox{\baselineskip0pt \lineskip-0.5pt
     \ialign{$\mathsurround0pt #1\hfil##\hfil$\crcr#2\crcr\sim\crcr}}}
\def\lsim{\mathrel{\mathpalette\oversim<}}    

\title[The PNe of the Sgr dSph galaxy]
{The  Planetary Nebula population of the Sagittarius Dwarf Spheroidal Galaxy}
\author[A.A. Zijlstra et al.]
       {Albert~A.~Zijlstra,$^1$ \thanks{E-mail: 
                               \tt a.zijlstra@manchester.ac.uk}
        K. Gesicki,$^2$   J.~R. Walsh$^3$,
       D. P\'equignot,$^4$  P.A.M. van Hoof,$^5$ 
       \newauthor  and D. Minniti$^6$ 
\\
        $^1$University of Manchester, 
          School of Physics \&\ Astronomy, P.O. Box 88, 
          Manchester M60 1QD, UK\\
        $^2$Centrum Astronomii UMK, ul.Gagarina 11, PL-87-100 Torun, Poland \\
        $^3$European Coordinating Facility, European Southern Observatory,
             Karl Schwarzschildstrasse 2, 85478 Garching, Germany \\
        $^4$Laboratoire d'Astrophysique Extragalacticque et de Cosmologie
             associ\'e au CNRS (UMR 8631) et \`a l'Universit\'e Paris 7,
             DAEC, \\ Observatoire de Paris-Meudon, F-92195 Meudon C\'edex, 
             France\\
        $^5$Royal Observatory of Belgium, Ringlaan 3, 1180 Brussels, Belgium\\
        $^6$Department of Astronomy, P. Universidad Cat{\'o}lica, 
             Casilla 306, Santiago 22, Chile \\
}
\pubyear{2006}

\begin{document}

\maketitle

\begin{abstract}
  The identification of two new Planetary Nebulae in the Sagittarius Dwarf
  Spheroidal Galaxy (Sgr) is presented. This brings the total number to four.
  Both new PNe were previously classified as Galactic objects.  The first,
  StWr 2-21, belongs to the main body of Sgr, from its velocity and location.
  The second, the halo PN BoBn\,1, has a location, distance and velocity in
  agreement with the leading tidal tail of Sgr. We estimate that 10 per cent
  of the Galactic halo consists of Sgr debris. The specific frequency of PNe
  indicates a total luminosity of Sgr, including its tidal tails, of $M_{\rm
    V}=-14.1$.  StWr\,2-21 shows a high abundance of [O/H]\,$=-0.23$, which
  confirms the high-metallicity population in Sgr uncovered by Bonaficio et
  al.  (2004).  The steep metallicity--age gradient in Sgr is due to ISM
  removal during the Galactic plane passages, ISM reformation due to stellar
  mass loss, and possibly accretion of metal-enriched gas from our Galaxy.
  The ISM re-formation rate of Sgr, from stellar mass loss, is $5 \times
  10^{-4} \rm \, M_\odot\,yr^{-1}$, amounting to $\sim 10^6 \rm \, M_\odot$
  per orbital period.
 
  {\it HST} images of three of the PNe reveal well-developed bipolar
  morphologies, and provide clear detections of the central stars. All three
  stars with deep spectra show WR-lines, suggesting that the progenitor mass
  and metallicity determines whether a PN central star develops a WR spectrum.
  One Sgr PN belongs to the class of IR-[WC] stars.  Expansion velocities are
  determined for three nebulae.  Comparison with hydrodynamical models
  indicates an initial density profile of $\rho \propto r^{-3}$. This is
  evidence for increasing mass-loss rates on the AGB. Peak mass-loss rates are
  indicated of $\sim 10^{-4}\rm \, M_\odot\, yr^{-1}$.
  
  The IR-[WC] PN, He\,2-436, provides the sole direct detection of dust in
  a dwarf spheroidal galaxy, to date.
\end{abstract}


\section{Introduction}

Planetary Nebulae (PNe) in Local Group dwarf galaxies provide valuable
indicators of the star formation history. They are especially important for
the age-metallicity relation, since they allow us to break the
age--metallicity degeneracy inherent in colour-magnitude diagrams.  Their
abundances can provide stellar nucleosynthesis information for metallicities
often very different from local, Galactic PNe. They are also important for
studying PN evolution itself. Knowing the age and metallicity of the stellar
population also allows one to study the formation and evolution of the PNe as
function of these properties. In particular, PNe form during the catastrophic
mass loss which terminates the Asymptotic Giant Branch (AGB) evolution. The
dependence of the mass loss on metallicity is not understood
\citep{Zijlstra2004,VanLoon2000,VanLoon2005}, and PNe in systems with
well-determined stellar populations are crucial for studying this important
problem.

The Sagittarius dwarf galaxy (Sgr) is our nearest surviving neighbour, at a
distance of 25\,kpc, less than half that of the Large Magellanic Cloud. Its
location behind the Galactic Bulge contributed to its late discovery
\citep{Ibata1994}.  The present mass and luminosity are low, within the range
of dwarf spheroidals.  The abundances range from SMC-like to LMC-like for
different star-formation epochs \citep[e.g.][]{Layden2000, Alard2001}.  Sgr is
strongly disrupted by its interaction with the Milky Way \citep{Putman2004,
  Helmi2001, DohmPalmer2001}, and was initially a substantial galaxy, in
accordance with its relatively high metalicity. Some distant carbon stars in
the Galactic halo may derive from the tidal streams of Sgr \citep{Mauron2005}.
This has also been suggested for distant halo globular clusters
\citep{Bellazzini2003}.  What fraction of the Galactic halo was strippped from
Sgr is not known.

\citet{Zijlstra1996} discovered that two previously catalogued (Galactic)
planetary nebulae were located in the Sagittarius dwarf galaxy on the basis
of their location and radial velocities.  They are thus the closest
extra-galactic PN. A detailed analysis based on ground-based spectra and radio
continuum data \citep{Dudziak2000} showed that the two nebulae have the same
stellar progenitor mass of 1.2M$_\odot$ and almost identical progenitor light
element abundances. One of the nebulae, Wray~16-423, underwent PN ejection
about 1500\,yr previous to its twin He~2-436.  A differential abundance
analysis revealed the first conclusive evidence for third-dredge-up oxygen
enrichment \citep{Pequignot2000}.  That the two PN have such closely matching
ages and abundances of non-dredged-up elements, pointed very strongly to their
belonging to the same, short-lived star-formation episode.

In this paper we present a new member of the Sgr PN population, arising from a
recent, higher-metallicity star-formation epoch.  One further Galactic halo PN
is also found to be a Sgr member, located in the leading tidal arm. This object
traces an older, much lower metallicity population.

Section 2 discusses the PNe membership of Sgr. In Section 3 we present new HST
images for three PNe, and deep spectroscopy of the new high-metallicity PN.
Kinematical models are discussed in Section 5. The central stars and the
nebular morphologies are discussed in Section 6. Abundances and 
implications for the evolution of the Sagittarius dwarf galaxy are discussed
in Section 7.

\section{Membership}
\label{association}

\begin{table*}
\caption[]{\label{location}
The Sgr PNe. For the first three objects, stellar magnitude, 
nebular flux, and diameters are
derived from the HST images presented here. Data for BoBn\,1 are from
\citet{Wright2005} but the position was redetermined. Stellar 
classification is discussed below.
 }
\begin{flushleft}
\begin{tabular}{ l c c c c l l l l l l}
\cline{1-10}
\noalign{\smallskip}
  PN         & l & b & RA & DEC &  $v_{\rm hel}$  
    & $\rm m_V$ (star) & $\log F(\rm H_\alpha$)
                       & diameter & star\\
             &   &   & \multicolumn{2}{c}{(J2000)}  & (km\,s$^{-1}$) 
             & (mag) & (erg cm$^{-2}$ s$^{-1}$) &  (arcsec)
\\
 \cline{1-10}
\noalign{\smallskip}
He\,2-436      & 4.8  & $-$22.7 & 19\ 32\ 06.70 & $-$34 12 57.4 & +133  & 
         $17.25\pm0.1$     & $-$11.74 &  0.6 & [WO\,4]/[WC\,4] \\
StWr\,2-21     & 5.2  & $-$18.6 & 19\ 14\ 23.35 & $-$32 34 16.6 & +129  & 
          $19.7\pm0.1$    & $-$12.26 & 2.7 & [WO\,2]\\
Wray\,16-423   & 6.8  & $-$19.8  & 19\ 22\ 10.63 & $-$31 30 38.7 & +133  
        & $18.85\pm0.2$     & $-$11.55 & 1.45 & [WC\,4]/wels \\
BoBn\,1       & 108.4 & $-$76.1 & 00 37 16.03 & $-$13 42 58.5 & +174    &
                           & $-$12.425   & 2 \\
\cline{1-10}        
\noalign{\smallskip}
\end{tabular}       
\end{flushleft}     
\end{table*}        

\subsection{A Sgr-core member: StWr\,2-21}

The main body of Sgr lies behind the Bulge, at a Galactic latitude of minus
15--20 degrees. The distance is between $24\pm2\,$kpc \citep{Alard1996} and
$d=26.3\pm1.8$\,kpc \citep{Monaco2004}.  Confusion with the Bulge foreground
population is important, but there are in fact relatively few Bulge PN more
than 15 degrees from the Galactic Centre.  The two known members, Wray\,16-243
and He\,2-436, were identified as located in the main body of Sgr based on
their identical radial velocities \citep{Zijlstra1996}.  Another object
located in the same region (Hb\,8) was shown to have a very different velocity
and was classified as a foreground object.  Further PNe in Sgr, wrongly
classified as Galactic Bulge PN, are a distinct possibility: many Bulge PNe
have no known velocity, and there are suggestions that the Sagittarius galaxy
could have had a total mass $>$10$^{9}$M$_\odot$ \citep{Jiang2000}. We
therefore conducted a search through the literature for catalogued PN in the
direction of the Sagittarius galaxy, and to test for a radial velocity
compatible with Sgr membership.

A little studied PN, StWr 2-21 (PN\,G005.2$-$18.6), was identified situated
within 2$^\circ$ of Wray\,16-423.  Its velocity was measured using a newly
obtained WHT spectrum (see Section \ref{spectra}).  The mean measured
heliocentric radial velocity of the H$\beta$ and [O~{\sc III}]\,4959,5007\AA\ 
lines was 129$\pm$1 kms$^{-1}$ (error on mean).  This agrees to within 5
km\,s$^{-1}$ with the other two Sgr PNe, leaving little doubt about the
association with Sgr.

All three identified members are located within the southern extension of Sgr,
in the same area where its globular clusters are found.  There is no PN
candidate in the core region of Sgr centred on M54 \citep[centre $l,b =
5.57,-14.17$, core radius 234 arcmin:][]{Majewski2003}.  The only
catalogued PN in this region, He 2-418, has an acceptable velocity of +115
km\,s$^{-1}$ but a diameter of 14 arcsec \citep{Ruffle2004} which is too large
for a bright PN at the distance of Sgr.  Assuming that this region has been
surveyed to uniform depth, the lack of bright PNe towards the assumed bound
core region of Sgr is noticable. There is no evidence of high extinction in
this direction.

\citet{Walsh1997} suggested that PRMG\,1 (PN\,G006.0$-$41.9) was a possible
Sagittarius PN. Its abundance is similar to those of
the two Sagittarius PNe \citep{Dudziak2000}. We can now
rule out this proposed association.  Observations with the NTT EMMI
spectrometer of this 4-arcsec nebula showed that its heliocentric velocity is
$+$20$\pm$2 kms$^{-1}$ (LSR velocity $+$24kms$^{-1}$). The Sagittarius 
stream 20 degrees from the main body is thought to have a velocity close to
100 km\,s$^{-1}$ \citep{Ibata2001} which does not favour membership.
(\citet{Helmi2001} predict velocities at this position closer to 50
km\,s$^{-1}$.) Recent mappings show that the tidal plane of Sgr does not pass
through this position \citep{Majewski2003}. We conclude that PRMG~1 is a
foreground object.

\subsection{A possible leading-tail member: BoBn 1}

We also searched along the plane of the tidal tails \citep{Majewski2003,
  Newberg2003}. This identified BoBn\,1, a well-studied halo PN, as a
candidate. Its heliocentric velocity of +174\,km\,s$^{-1}$ \citep{Wright2005}
is opposite to the trailing tail at this location \citep{Majewski2004}.
However, the positional agreement is sufficiently close (and halo PNe are
sufficiently rare) that the question may be asked whether BoBn\,1 could be
part of a separate tail.

A dynamical simulation by \citet{Law2005} shows the various streams of debris,
separated into trailing and leading tails, and associated with different tidal
dislocations events. The streams are always located in the same orbital plane
but are moving in different directions and are at different distances. BoBn\,1
is located in a direction where the leading and trailing streams cross.  At
this location the leading stream is moving outwards, along the line of sight.
The particular velocity is model dependent, but the velocity of BoBn\,1 is
within the range of the predictions. The heliocentric distance to the leading
stream at this location is around 20\,kpc. The distance to BoBn~1 is quoted as
22.5\,kpc \citep{Hawley1978} and 16.5\,kpc \citep{Henry2004}. (Distances to
PNe are difficult to ascertain, and the quoted distance is uncertain by at
least 30 per cent.) The agreement in position, velocity and distance makes an
association of BoBn\,1 and the Sgr leading tail a likely one.

The leading tail is observationally not as well constrained as the trailing
tail \citep{Law2005}.  The identification of further stars in the Sgr leading
tail will be needed to remove the model dependence in the proposed
association.The accurate velocity of BoBn\,1 may also be useful in
constraining the Sgr tidal models.

No other catalogued PNe were identified as potential members. Other newly
discovered PNe in the Sgr core region \citep{Parker2003} are extended and
faint and certainly not Sgr members.  The list of possible PNe in the
ESO-Strasbourg catalogue of Planetary Nebulae \citep{Acker1992} contains three
further objects of interest (K\,2-4, Y-C\,43, ESO\,340-34), but all are
nebulae of large radius. They are possibly misclassified galaxies.

The four identified Sgr PNe are listed in Table \ref{location}.

\section{HST imaging}

We obtained SNAPshot WFPC2 images in three different filters of the three PNe
in the main body of Sgr.  We used a sequence of two
cosmic-ray-split H$\alpha$ exposures, one V-band exposure and two [O\,{\sc
  iii}] exposures. For He2-436 and Wray 16-423, exposures times are 100
seconds for each H$\alpha$ exposure, 60 sec for the V-band and 80
seconds for each [O\,{\sc iii}] exposure.  The faintest object, StWr\,2-21,
was observed for 2 times 140 sec in H$\alpha$ and [O\,{\sc iii}].  The objects
were placed on the CCD of the Planetary Camera (PC).  The pixel scale is
$0.0455^{\prime\prime}$.  Pipe-line reduced images were retrieved.

The calibration of the three filters used (F502N, F547M, F656N) is described
in \citet{ODell1999}.  The filter response is taken from the WFPC2 handbook.
The F656N filter is centred on H$\alpha$ wth a peak transmission above 70\%.
It has a $\sim 10\%$ response at the [N\,{\sc ii}] 6548\AA\ line and a few per
cent at the 6584\AA [N\,{\sc ii}] line. The He\,{\sc i} line at 6560\AA\ can
also contribute but this line is weak in all cases.  The F502N filter has a
60\%\ transmission at the [O\,{\sc iii}] line without significant pollution
from other lines. The F547M is used as the continuum filter, and is best
suited to detect the central star, and other possible confusing stars--it is
little affected by strong emission lines. Nebular bound-free continuum
dominates the nebular flux in this filter.

Photometric calibration was done with the calibration constants (PHOTFLAM and
the Vega-based zeropoint) listed in the WFPC2 handbook.  The central stars of
all three nebulae are detected in the F547M images. The H$\alpha$ image was
scaled and subtracted from the F547 image, to remove the nebular continuum.
The stellar magnitudes, converted to Johnson V, are listed in Table
\ref{location}.  They were measured in a circular apertures of 7-11 pixels
radius, using aperture corrections shown in the WFPC2 handbook for the V-band
filter and the PC.  The accuracy is 0.1--0.2 mag, limited by the faintness of
the stars.  In addition to the
magnitudes, Table \ref{location} also lists the line fluxes measured from the
images.

\begin{figure*}
\includegraphics[angle=270, width=17cm]{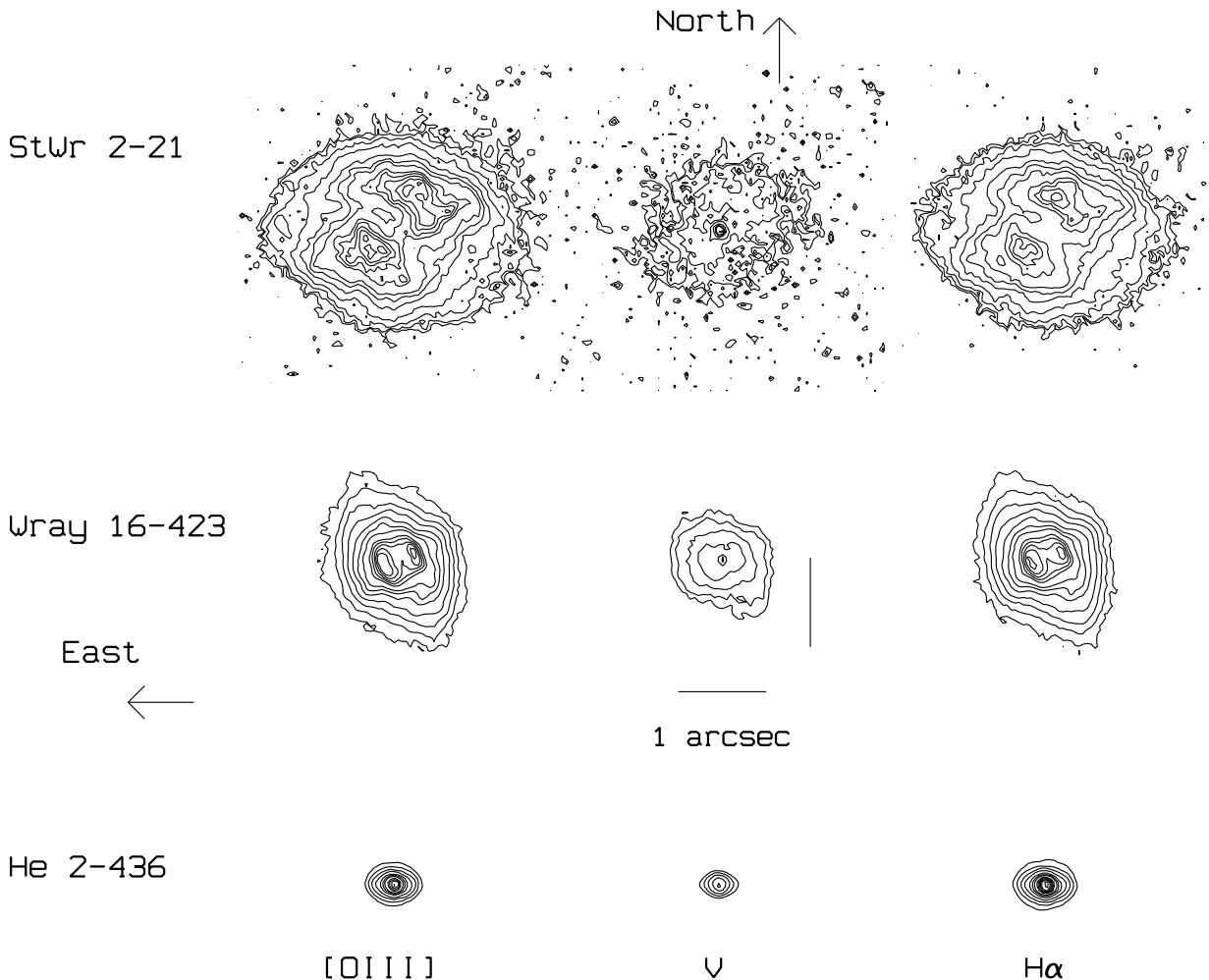}
\caption{\label{contours} Contour plots in [O\,{\sc iii}], V-band and 
  H$\alpha$ of the three Sgr PNe. The peak fluxes are scaled to give
  approximately the same peak value for all [O\,{\sc iii}] images. A small
  constant was subtracted from the StWr 2-21 images to suppress the lowest
  contour. Contour levels are at 0.5,1,2,4,8,12,16,24,28,32,36,40 scaled
  counts. At the distance of Sgr, 1 arcsec corresponds to $3.9 \,
  10^{17}\,$cm}
\end{figure*}

Fig. \ref{contours} shows the resulting images. Here the images were rotated
to the same orientation, while rebinning to keep the same square pixel size.
All images are shown to the same spatial scale.  The images reveal bipolar
structures in all three PNe.  The large difference in size is clear.  The peak
intensity drops by a factor of roughly 100 between He\,2-436 and StWr\,2-21
(in the depicted images the fluxes have been scaled to the same peak for the
[O III] images).  He\,2-436 is very compact. The inner radius is only just
resolved, as shown by the two peaks separated by a single pixel.  It is
significantly elongated. Wray\,16-423 shows a complicated morphology, with an
inner bipolar structure surrounded by a second, fainter structure rotated at
an angle relative to the inner countours.  The outer contours show elongation.
Finally, StWr\,2-21 is the largest nebula, and shows an elliptical outer
structure and a bipolar inner structure.  There is some evidence for a
wind-blown bubble towards the East which in [O\,{\sc iii}] appears as a closed
outer shell. As for Wray\,16-423, a rotation between the inner and outer
structures is apparent.

\begin{figure*}
\includegraphics[width=17cm, height=12cm]{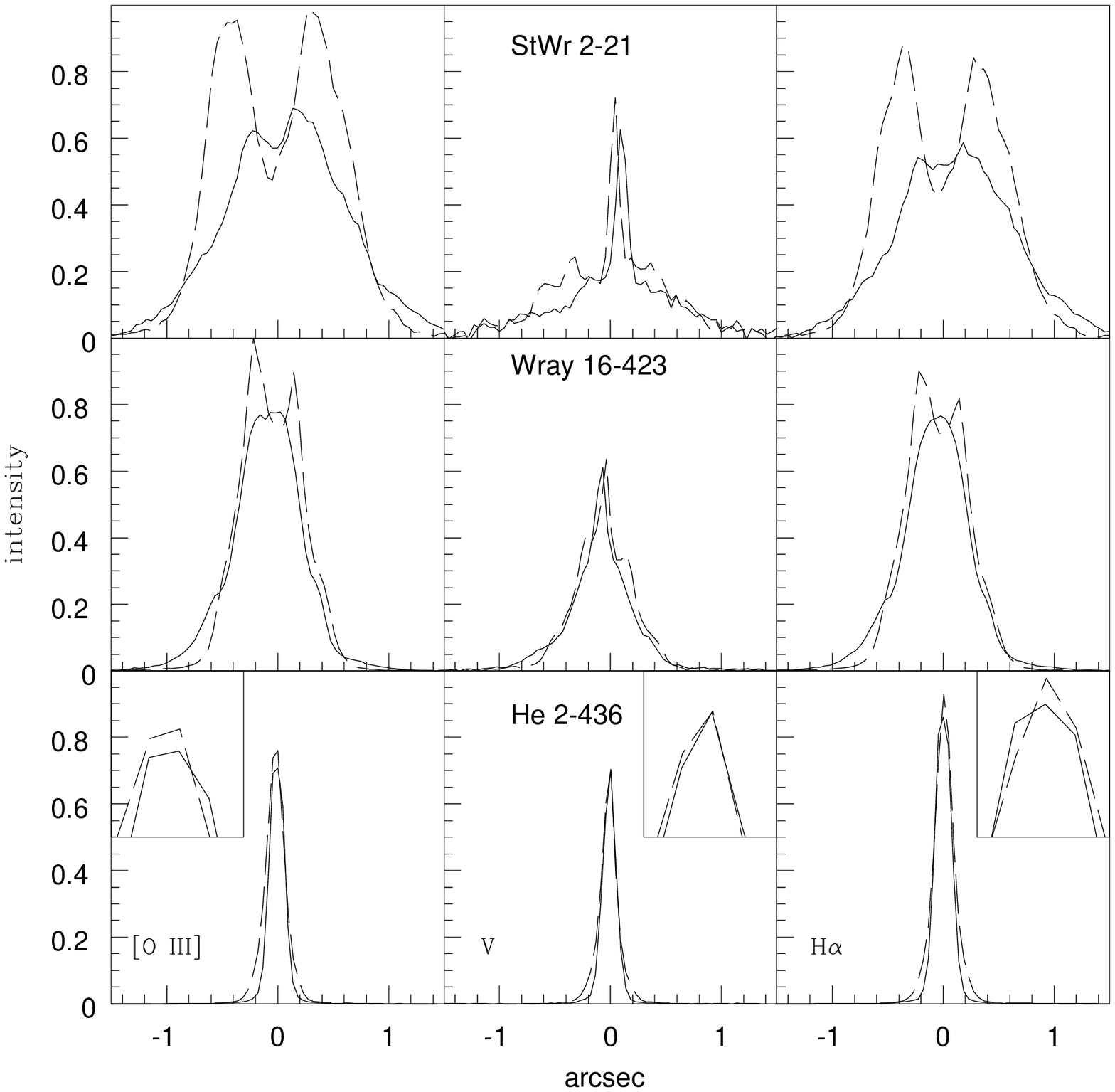}
\caption{\label{profiles} Cross sections of the three nebulae observed with 
  HST. The profiles are summed over three adjacent rows.  Dashed lines are
  profiles along the minor axis (typically showing the two peaks) and
  continuous lines are along the major axis. The insets show the inner 0.2
  arcsec of He 2-436, showing the flattening in [O\,{\sc iii}] and H$\alpha$,
  compared with the central peak seen in V which is due to the central star.
}
\end{figure*}

Fig. \ref{profiles} shows radial intensity profiles along the minor axis and
along the major axis (90 degrees rotated). These were obtained by rotating (and
rebinning to the original square pixel size) to align the major axis with
the image axes, and averaging three adjacent rows or columns through the
geometric centroid of the source.  The V-band profiles show the central star,
although in the case of He2-436 only marginally.

\begin{table}
\centering
\caption[]{StWr\,2-21 Emission-line flux ratios as measured from the
 low-dispersion spectra, 
 relative to I(H$\beta$)=100.0..} 
\label{low-disp}
\begin{tabular}{llrrrr}
Rest $\lambda$ & Identification & Obs. flux & error 
   &  Obs. flux   \\
    (\AA)      &                &  (EFOSC2) & &  (ISIS) \\
\hline
     3425 & [Ne~V]   &           1.73  &      0.23 &       \\
     3440 & O~III    &          10.16  &      0.19 \\
     3705 & H~I      &           0.59  &      0.08 \\
     3726 & [O~II]   &          34.36  &      0.16 &       \\
     3753 & O~III    &           3.90  &      0.07 \\
     3769 & H~I      &           3.31  &      0.09 \\
     3796 & H~I      &           4.70  &      0.07 \\
     3814 & He~II    &           1.51  &      0.06 \\
     3835 & H~I      &           6.67  &      0.09 \\
     3869 & [Ne~III] &          70.97  &      0.20 &       \\
     3889 & H~I +He~I &         16.69  &      0.11 &       \\
     3922 & He~II    &           0.40  &      0.11 \\
     3969 &  H~I + [Ne~III] &   35.89  &      0.15 &       \\
     4026 & He~I     &           2.05  &      0.14  &       \\
     4070 & [S~II] (blend)   &           1.99  &      0.08  \\
     4101 & H$\delta$  &        24.60  &      0.11  &       \\
     4188 & C~III    &           0.27  &      0.05  &       \\
     4200 & He~II    &           0.61  &      0.04 \\
     4267 & C~II     &           0.78  &      0.08  &       \\
     4340 & H$\gamma$ &         43.710 &      0.11  &       \\
     4363 & [O~III]  &          10.32  &      0.07  &       \\
     4387 & He~I     &           0.39  &      0.10   &      \\
     4471 & He~I     &           3.40  &      0.07   &      \\
     4543 & He~II    &           1.22  &      0.07 \\
     4648 & C~III    &           2.75  &      0.57    &       \\
     4685.7 & He~II    &          41.86  &      0.11  &  38.1 \\
     4711.4 & [Ar~IV] + He~I &     4.05  &      0.06  &   4.0 \\
     4740.2 & [Ar~IV]  &           3.15  &      0.07  &   3.3 \\
     4861.3 & H$\beta$ &         100.00  &      0.00  & 100.0 \\
     4923 & He~I     &           1.23  &      0.06    &       \\
     4958.9 & [O~III]  &         355.65  &      0.51  & 359.4  \\
     5006.8 & [O~III]  &        1062.02  &    189.91  & 1047.2 \\
     5412 & He~II    &          3.69   &      0.04 \\
     5518 & [Cl~III]  &         0.50   &      0.04 &     \\
     5539 & [Cl~III]  &         0.46   &      0.04 &     \\
     5755 & [N~II]    &         0.30   &      0.11 &     \\
     5805 & C~IV      &         0.90   &      0.11 \\
     5876 & He~I      &        11.04   &      0.13 &      \\
     6300 & [O~I]     &         1.37   &      0.07 &      \\
     6312 & [S~III]   &         2.12   &      0.04 &     \\
     6363 & [O~I]     &         0.46   &      0.06 \\
     6408 &  He~II    &         0.19   &      0.04 \\
     6548 & [N~II]    &         2.01   &      0.06 \\
     6563 & H$\alpha$  &      279.07   &     49.76 &     \\
     6584 & [N~II]    &        11.53   &      0.07 &     \\
     6678 & He~I      &         2.65   &      0.04 &     \\
     6716 & [S~II]    &         2.00   &      0.04 &     \\
     6731 &  [S~II]   &         2.97   &      0.04 &     \\
     6891 & He~II     &         0.35   &      0.04 \\
     7065 & He~I      &         2.33   &      0.07  &    \\
     7136 & [Ar~III]  &         9.36   &      0.07  &    \\
     7175 & He~I      &         0.57   &      0.06 \\
     7211 &           &         2.98   &      0.09 \\
     7322 & [O~II]    &         2.26   &      0.08 &   \\
     7362 &           &         0.43   &      0.07 \\
\hline
\end{tabular} \\
\begin{list}{}{}
\item Observed H$\beta$ flux = 1.29 $\pm$ 0.01
$\times$10$^{-13}$  (EFOSC2) and  1.06 $\times$10$^{-13}$ (ISIS) ergs 
cm$^{-2}$ s$^{-1}$
\end{list}
\end{table}

\section{Spectroscopic observations}

\subsection{Low resolution spectroscopy of StWr\,2-21}
\label{spectra}

We observed StWr\,2-21 with the ISIS spectrometer on the 4.2m WHT La Palma
telescope on 1999 July 19. The holographic H2400B grating was used, and the
EEV12 CCD detector with a pixel scale of 0.11\AA\ and a coverage of 445\AA. A
slit width of 1.5\,arcsec was employed giving a line width of 5 pixels and a
resulting spectral resolution of 0.53\AA. A single exposure of 900s was
obtained with the slit at a parallactic angle; the seeing was measured at
about 2\,arcsec. The spectrum was centred at 4800\AA\ allowing the He~II
4686\AA, H$\beta$ and the [O\,{\sc iii}]4959,5007\AA\ lines to be studied.  A
Cu-Ar arc lamp exposure was employed for wavelength calibration and a
5th-order fit to the line positions was employed (r.m.s. on the fit
$<$0.01\AA).  An exposure of the spectrophotometric standard star
BD+33$^\circ$2642 \citep{Oke1990} was made with a broad (5.0\,arcsec) slit and
used to establish the absolute flux calibration for the StWr\,2-21 spectrum.

A deep, low-resolution spectrum with larger wavelength coverage was obtained
with EFOSC2 on the ESO 3.6m telescope \citep{Patat1990}.  An acquisition image
was obtained with an [O~III] filter (centre 5004\AA, width 56\AA; ESO Filter
\#689) and spectra at a parallactic angle with a 1.2\,arcsec slit were obtained
over the wavelength ranges 3050-6100 (grism \#03) and 4090-7520\AA\ (grism
\#04).  The seeing was 1.6\,arcsec measured from the acquisition image.  The
exposure time was 1800s, split into two sub-exposures; separate 300s exposures
were also obtained to provide all strong lines unsaturated. The measured
spectral resolution was 11\AA\ for the bluer spectrum and 12\AA\ for the
redder spectrum. Broad-slit observations of the standard star EG274
\citep{Oke1990} were obtained with both gratings for flux calibration.
 
The spectral images were bias-subtracted and flat-fielded by exposures to a
continuum lamp illuminating the inside of the dome. Wavelength calibration was
applied using the exposures to a Cu-Ar arc lamp fit by a third order
polynomial.  The spectrum of the PN was extracted over the extent of the
emission, sky subtracted and flux calibration applied from the spectra of
EG274. Table \ref{low-disp} presents the observed line fluxes determined by
interactively fitting Gaussians to the lines for the longer exposures, with
substitution of the fluxes for the saturated lines from the shorter exposures
([O~III] doublet and H$\alpha$).  Errors on the line fluxes, from propagation
of the photon counts on the lines and the error in fitting the underlying
continuum, are also listed in the Table.

The extinction is zero within the errors, as derived from the hydrogen line
ratios. For comparison, the extinction towards Wray\,16-423 is $E(B-V)=0.14$
\citep{Walsh1997}, consistent with the forground extinction towards Sgr.
He\,2-436 has higher extinction due to dust within the nebula. 

\subsection{Echelle observations}

He 2-436 and StWr 2-21 were observed with the ESO NTT telescope during June
2001, as part of a larger survey of PNe towards the Galactic Bulge
\citep{Gesicki2006}.  The echelle spectra were obtained using the red arm of
the combined imager/spectrograph EMMI. Echelle grating 14 was used with cross
disperser 3.  The wavelength coverage is approximately 4300--8100\AA.  The
slit width was 1\arcsec, providing a resolution of 60\,000.  The slit length
is limited to 3\,arcsec by the need to avoid order overlap.  The seeing (worse
than 1 arcsec) did not allow us to use the spatial resolution inherent in the
data, and we summed the spectrum over the slit length.

Wavelength calibration was done using a ThAr lamp. The data was flat-fielded
and corrected for the response function using a standard star. No absolute
flux calibration was attempted. The exposure time of 120 sec aimed at
obtaining good line profiles for the strongest lines in the spectrum only.
Line profiles used to obtain velocity fields were determined for four lines.
For Wray 16-423, similar but much deeper observations were published in
\citet{Gesicki2003a}. The data reduction used here is as described in that
paper.

The line ratios from these data are listed in Table~\ref{low-disp}.
In the following sections we describe models based on these new data. These
are combined with existing models for the previously known Sgr PNe. No new
data was obtained for BoBn\,1, and this object is  not included.

\section{Expansion velocities}
\label{velocities}

\subsection{Description of the models}

The internal kinematics of PNe are dominated by expansion. The expansion
velocity together with the physical size of the nebula gives a measure
of the elapsed time since the ejection of the nebula, the dynamical age.
This can be compared to the thermal age, defined by the temperature of the
central star and its rate of temperature increase.  The latter is a sensitive
function of the mass of the central star. The dynamical age together with
the stellar temperature can therefore be used to derive the stellar mass.

The internal kinematics are dominated by the expansion of the nebulae.
Hydrodyamical models \citep[e.g.][]{Schoenberner2005} predict that the
expansion velocity is not uniform, but that different layers (at different
radii) expand at different rates. The echelle spectra clearly show the
broadened lines due to the expansion, but determination of the internal
velocity field is needed to define an overal expansion velocity. The
mass-weighted expansion velocity \citep{Gesicki2000} is used for calculating a
dynamical age.

\begin{figure*}
\includegraphics[width=17cm, clip=]{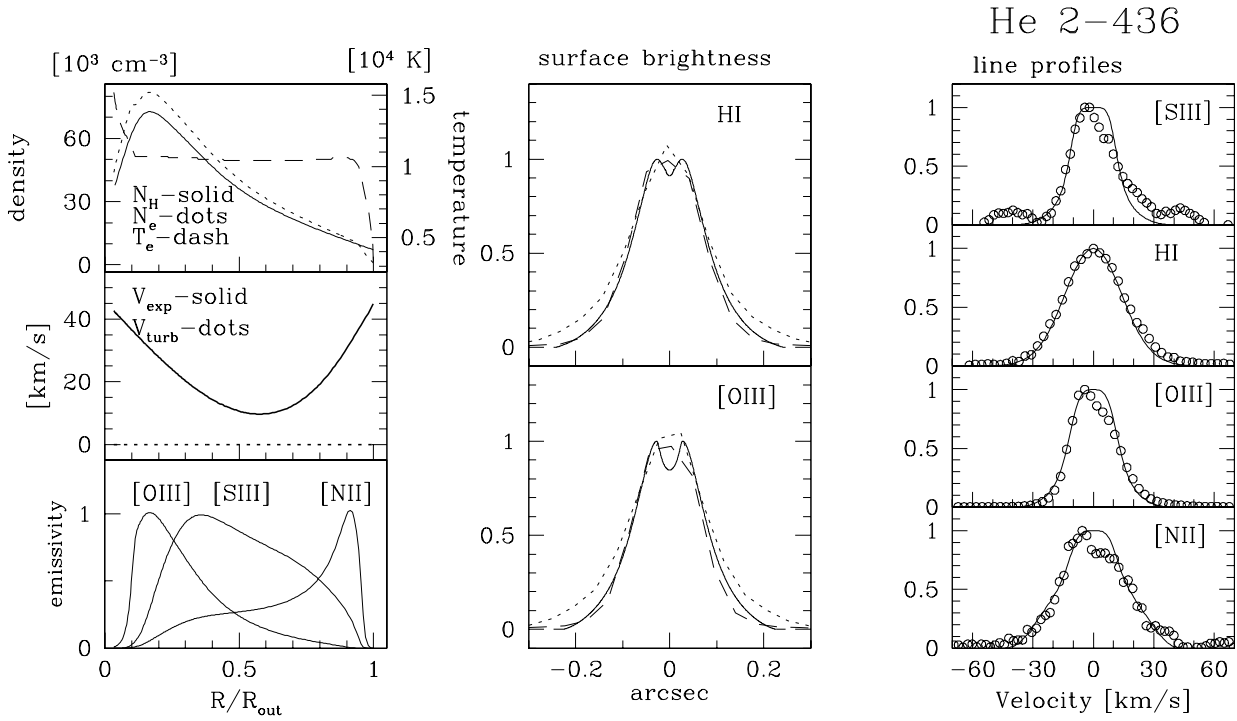}
\caption{\label{model_h} Model fits to the line profiles of He 2-436. Left:
parameters of the ionization model, showing the radial distribution of the
including the density, electron temperature, velocity field, and the ionic
distribution. Middle: radial surface brightness profiles: model (solid lines)
versus the observed major and minor axis profiles from the HST images,
fig. ref{profiles}. Right: Fitted line profiles. }
\end{figure*}

\begin{figure*}
\includegraphics[width=17cm, clip=]{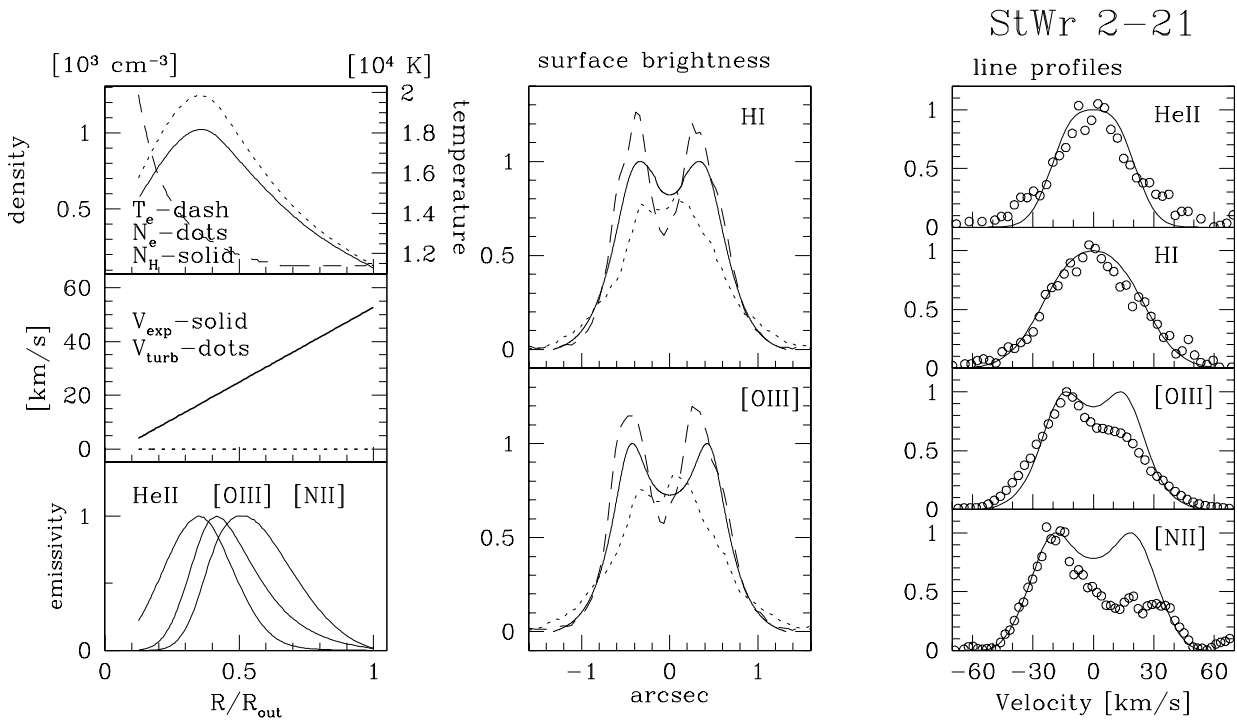}
\caption{\label{model_s} Model fits to the line profiles of StWr 2-21.  
details as in Fig. \ref{model_h}. }
\end{figure*}

We will follow the usual procedure, as outlined in \citet{Gesicki2003a}. Here
we first use a photoionization model to determine where within the nebula a
certain line originates, and use the measured line profiles to fit a velocity
field as function of radius. The model assumes a spherical symmetric nebula.
The models we use (the Torun models) do not currently implement asymmetric
nebulae, and neither do we have sufficient observational constraints to fit
such models. (For instance, asymmetric velocity models require knowledge of
the angle between the polar axis and the plane of the sky, which would require
spatially resolved echelle spectra. Our spectra lack spatial information.)
The Torun models also approximate the star with a black body.  However, the
resulting expansion velocity is found to be relatively robust to these
assumptions.

The computer model first solves for the photo-ionization equilibrium, where we
attempt to reproduce the observed line intensities and the observed radial
surface brightness distribution. (This photo-ionization model is by necessity
different from that used in section \ref{abundance_model}, and is not suitable
for abundance determinations.)  We compare the models with both major and
minor axis cross sections of the nebulae, but do not attempt to fit these
individually. The line profiles are calculated, taking into account the slit
aperture (which may not cover the full nebula) and the seeing (which scatters
light rays from outside the aperture into the slit).  The velocity field is
adjusted to fit the line profiles of all lines simultaneously.  In the present
analysis the computer codes are adapted to work with an optimization routine.
We applied the publicly available ``PIKAIA'' computer code which is based on a
genetic algorithm \citep{Gesicki2006}. The line profiles include thermal
broadening, organized (expansion) velocities, and turbulence. However, for the
modelled Sgr PNe we did not find evidence for a turbulent component.

\subsection{Kinematical models for three PNe}

\subsubsection{He 2-436}      
%

The spectra are spatially unresolved, as shown by the fact that the lines are
unsplit (Fig. \ref{model_h}). Four lines are suitable for the velocity
analysis; the weakest of these ([\ion{S}{iii}] 6312\AA) was filtered to smooth
the noise. The lines are approximately of Gaussian shape, with [\ion{O}{iii}]
and [\ion{S}{iii}] slightly narrower than the other two.

In accordance with the HST images we  build an appropriate
photoionization model assuming the density distribution is concentrated
towards the inner radius.  The chemical composition and line ratios have
been adopted from \citet{Dudziak2000} and \citet{Walsh1997}.
A parabola-like expansion velocity resulted in acceptable fit to the
observed line profiles.

The adopted solution is shown in Fig. \ref{model_h}.  Panels on the left
present the model structure.  Shown are: the assumed radial hydrogen density
distribution together with the computed electron density and temperature; the
derived velocity distribution; and the radial distributions of emissivities
for lines shown in right panels, normalized to unity.  The other panels
compare models with observations. In the middle the observed HST surface
brightness profiles from Fig.\,\ref{profiles} are compared to the computed
ones.  The observations (dots and dashes) are rescaled to the normalized model
profiles (draw line). The panels on the right compare the computed (solid
lines) with the observed (circles) emission line profiles.

%
%
\subsubsection{StWr 2-21}

The 2-arcsec nebula is expected to be partially spatially resolved by the
spectrograph slit. Only four emission lines are strong enough to be considered
in our analysis. The \ion{He}{ii} and H$\beta$ lines are unsplit,
[\ion{O}{iii}] shows a partial split, and [\ion{N}{ii}] is split more clearly
but is also more asymmetric. For the model fitting we used only the stronger
(blue) peaks from the asymmetric couple.

The best-fit model that reproduces both the H$\beta$ flux and the line
ratios of Table \ref{low-disp} is a density-bounded one. Such a model
reproduces the very similar size of the images in the two different
filters (Fig. \ref{contours}), because in this case the ionization
stratification is very weak and the emissivity distribution of
[\ion{O}{iii}] and H$\alpha$ follows the same density distribution. (The
computed emisivities show for \ion{He}{ii} and [\ion{N}{ii}] a clear
stratification, but no images are available to test this.)  To reproduce
the small size of the central cavity, we needed to apply a density
distribution concentrated towards the centre with inner radius close to
the central star.

The linearly increasing velocity field reproduces relatively well the
four observed emission lines.

\begin{figure*}
\includegraphics[width=17cm, clip=]{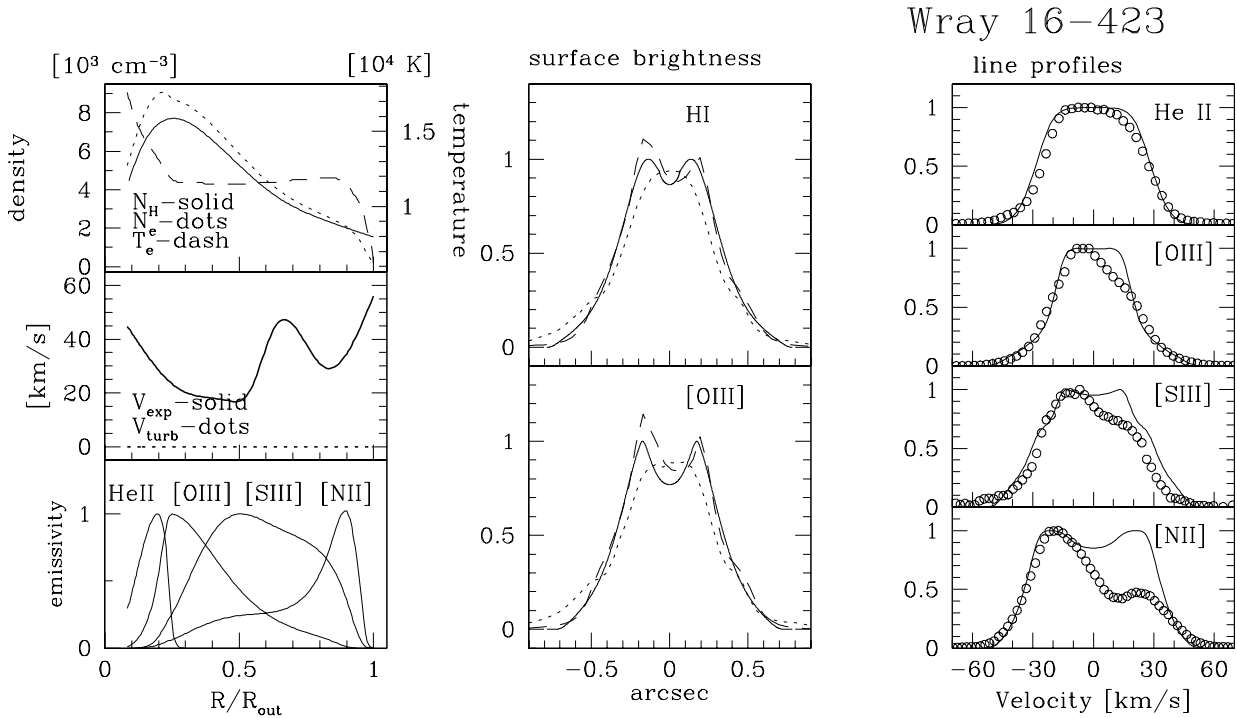}
\caption{\label{model_w} Model fits to the line profiles of Wray 16-423.  
details as in Fig. \ref{model_h}. }
\end{figure*}

%
%
\subsubsection{Wray 16-423}

An analysis of the velocity field of the nebula Wray\,16-423 was
published in \citet{Gesicki2003a}. The high quality spectrum allowed for
extraction of nine emission lines which strongly constrained the deduced
velocity field.  The result indicated a high velocity spike located
inside the nebula and also suggested a presence of a low density
extended outer region.

The previous modeling was done without the benefit of a high-resolution image.
We redid the modeling, searching for a density distribution that reproduces as
well as possible the new HST images. The obtained density field is similar to
that from \citet{Gesicki2003a}, but the maximum is placed somewhat further
from the centre and the outer density is higher.

With the improved density structure we returned to the old spectra.  The
genetic algorithm found a velocity field similar to that from
\citet{Gesicki2003a} with a high velocity region inside the nebula. The fit
considered eight of the nine lines shown in \citet{Gesicki2003a} (excluding
the most noisy one) but in Fig. \ref{model_w} we present only the most
representative four lines.  The velocity profile is more complicated than
found for the other two PNe, but this is partly caused by the fact that more
lines are available and that therefore the velocities as function of
emissivities are better constrained.  This allowed us to search for a
higher-order solution. The low-order behaviour is similar to the other
objects.

\subsubsection{Results}

Table \ref{velocityfields} presents the summary of the parameters of
Torun models for the three PNe. The important parameter are the
mass-weighted expansion velocity, $V_{\rm aver}$, the dynamical age,
and the core mass for which the dynamical and stellar time scales are
equal.

\begin{table}
\begin{flushleft}
\caption[]{\label{velocityfields} Parameters and results of the velocity 
field analysis}
\begin{tabular}{llll}
\cline{1-4}
\noalign{\smallskip}
                      & He 2-436 & Wray 16-423 & StWr 2-21 \\

\noalign{\smallskip}
\cline{1-4}
\noalign{\smallskip}

log F(H$\beta$)       & $-$11.4         &  $-$11.89   &  $-$12.6         \\

[O/H]                 & $-$0.52         & $-$0.51    &  $-$0.35      \\

$T_{\rm eff}$             & 58100       &  82000   &  146000    \\

$\log T_{\rm eff}$        & 4.76        &  4.92      &  5.16     \\

$\log L/L_{\odot}$       & 3.89         &  3.6     &  2.95         \\

model $R_{\rm out}$ [pc]  & 0.03          &  0.095     &  0.16        \\

model $R_{\rm inner}$ [pc]  & 0.001        &  0.008      &  0.02        \\

model $M_{\rm ion}$       & 0.084           &  0.40     &  0.25      \\

outer boundary        & ioniz        &  ioniz     &  density      \\

angular diam [arcsec] & 0.5             &  1.6    &  2.6        \\

$V_{\rm aver}$ [km/s]     & 19           &  32      &  32          \\

turbulence [km/s]     &  0           &  0      &  0           \\

dyn. age [yr]         & 1650         &  3100       &  5200           \\

dyn. core mass        & 0.62          &  0.61       &  0.61          \\

\noalign{\smallskip}
\cline{1-4}

\end{tabular}
\end{flushleft}     
\end{table}        

As expected, there is an age progression with He\,2-436 being the youngest and
StWr\,2-21 the oldest nebula. The difference is smaller than the size would
imply, as He\,2-436 has a small expansion velocity. The derived velocities are
within the usual range for Galactic objects \citep{Gesicki2000}. Available
data suggests that the metallicity has little effect on the PN expansion
\citep{Gesicki2003a}. This is in contrast to AGB outflow velocities which are
strongly dependent on metallicity \citep{Marshall2004}. The reason is that the
PN expansion is largely dominated by the overpressure within the ionized
region, and has accelerated significantly during the post-AGB phase. The
kinematic models show that all three stars have comparable core (final)
masses. There is a systematic uncertainty, in that some assumptions are made
on how the nebula accelerates, and that the radius used for the dynamical age
is taken as 0.8 of the outer (model) radius. (The latter because this is
approximately the mass-weighted radius.)

All three nebulae show an asymmetry in the line profiles with the blue
component stronger. The difference is very minor in He\,2-436, larger in
Wray\,16-423 and strongest in StWr\,2-21. In all cases do the ionization
levels tracing the outer-most regions (such as [O\,{\sc i}] and
 [N\,{\sc ii}]) show
the largest effect. The effect of the asymmetry on the line modelling is
discussed in \citet{Gesicki2003a}: the velocity structures are more regular 
than the density and/or temperature profile, as shown by the line profiles.
The cause for the blue asymmetry is not known, but one can speculate about
a relation to the movement of the nebulae through the gas in the Galactic
halo.

\section{Discussion: Stars and nebulae}
\label{abundance_model}

\subsection{Photo-ionization modelling}
\label{photomodel}

Emission lines are sensitive to the local conditions in the nebulae:
density, temperature, radiation field, and elemental abundances. This
sensitivity allows for accurate models, with improving accuracy when more
lines are available.  Deep spectroscopic observations of Wray\,16-423 and
He\,2-436 have previously been used to obtain photo-ionization models of both
nebulae \citep{Dudziak2000}.  Neither object could be well described by a set
of unique conditions, but each required two sectors with different covering
factors and conditions. Their model parameters are listed in
Table~\ref{models}.  The models have filling factor unity and assume constant
pressure within each zone.  Table~\ref{models} also lists the radii and
thickness derived from the HST images. The models were derived before the HST
images became available: the values predicted by the models  are largely
confirmed by the new images presented above. The structures
observed by HST confirm the need for at least two-zone models.  The
complexity in the images exceeds what can be reproduced in the models,
but the large-scale geometry confirms the predictive capability of the
models of  \citet{Dudziak2000}.

Photoionization modelling of StWr 2-21, again using NEBU, is discussed in
detail in P\'equignot et al. (2006, hereafter Paper~II). Models are based on
the fluxes listed in Table~\ref{low-disp}, but with  minor revisions
suggested by a spectrum synthesis.  The synthesis was used to investigate the
contribution of weak, not individually detected, lines on the spectrum. It
also allowed us to resolve blends, important at the moderate resolution of the
observations, and to detect Wolf-Rayet features (see Section~\ref{wr_lines}).

As for the two objects modelled by \citet{Dudziak2000}, at least two sectors
are necessary to reproduce the observed line flux ratios. With the constraint
of the nebular size derived from the HST images, the two-sector models for
StWr\,2-21 turn out to have a combined covering factor much less than unity.
A low-density gas with no specific signature in the spectrum may fill in most
of the 'empty' sector, as confirmed by trial calculations. Given that the
overwelmingly dominant sector is strongly matter-bounded, the existence of
such a component is not too surprising. Although it may not emit much of the
line flux, this sector may encompass a very significant fraction of the
ionized mass, which is therefore only a lower limit in the two-sector model.

The question of the uniqueness of the solution is discussed in Paper~II. It is
difficult to obtain unique models in the case of StWr 2-21. The carbon
abundance, which controls the energy output of the nebula and therefore the
effective temperature of the central star, is poorly constrained for lack of
UV spectra. Most optical recombination lines of carbon have mildly accurate
fluxes.  Moreover, following a suggestion made for other PNe, these
recombination lines may partly trace cool H-poor clumps embedded in the nebula
\citep[e.g.][]{Liu2004, Pequignot2003}.  The NEBU model selected in
Table~\ref{models} is therefore typical but not unique.  Nonetheless, the
oxygen abundance is relatively well defined.  Abundances derived from this
model are provided in Table~\ref{abundance-comp} and discussed in
Section~\ref{section_abundances}.

\begin{table*}
\caption[]{\label{models} Results of the multi-sector
photo-ionization models. Values for StWr\,2-21 are from Paper~II.
Values for He\,2-436 and Wray\,16-423 are from \citet{Dudziak2000}
Observed radii (third and fourth rows) are from HST data presented here 
(except for BoBn 1).
The covering factors are the fractions of the total solid angle, as seen from 
the central star, occupied by each model sector. The penultimate row gives 
the radial optical thickness at the ionization threshold of H where required.
The last row gives the fraction of the total ionized mass in each sector.}
\begin{flushleft}
\begin{tabular}{ l l l  l l l l l l l }
\cline{1-10}
\noalign{\smallskip}
              & \multicolumn{2}{c}{He 2-436} 
               & \multicolumn{2}{c}{Wray 16-423}
                & \multicolumn{2}{c}{StWr 2-21} 
                 & \multicolumn{2}{c}{BoBn 1} &  (units) \\
\cline{1-10}
\noalign{\smallskip}
 $T_{\rm eff}$ &    \multicolumn{2}{c}{$0.70\,10^5$}
               &    \multicolumn{2}{c}{$1.07\,10^5$}    
                &    \multicolumn{2}{c}{$1.19\,10^5$}
                 &    \multicolumn{2}{c}{$1.30\,10^5$}   & K \\
 $L$           &    \multicolumn{2}{c}{$5.40\,10^3$}       
               &    \multicolumn{2}{c}{$4.35\,10^3$}       
                &    \multicolumn{2}{c}{$4.10\,10^3$}
                 &    \multicolumn{2}{c}{$5.20\,10^3$}  
                                                        & L$_\odot$ \\
 $M_{\rm ion}$ &   \multicolumn{2}{c}{$0.02$}   
                &    \multicolumn{2}{c}{$0.25$}   
                  &    \multicolumn{2}{c}{$0.06$}
                   &    \multicolumn{2}{c}{$0.27$}  & M$_\odot$ \\
 $R_{\rm inner}$ (HST, observed)  &   \multicolumn{2}{c}{$6\,10^{-3}$:} 
                      &   \multicolumn{2}{c}{$2\,10^{-2}$} 
                      &   \multicolumn{2}{c}{$4\,10^{-2}$}
                      &   \multicolumn{2}{c}{--} & pc \\
 $\Delta R$  (HST, observed)    &   \multicolumn{2}{c}{$3\,10^{-2}$} 
                      &   \multicolumn{2}{c}{$5\,10^{-2}$} 
                      &   \multicolumn{2}{c}{$8\,10^{-2}$}
                      &   \multicolumn{2}{c}{($\sim$0.1)} & pc \\
\cline{1-10}
\noalign{\smallskip}
 Model sectors &  I & II &  I & II & I & II & I & II \\
 Ionization/density bounded &   ionization & ionization 
                 & ionization & density
                  & ionization & density
                   & ionization & density \\
 $<\!N_{\rm H}\!>$   &   $2.7\,10^5$   &   $2.0\,10^4$     
              &  $9.5\,10^4$  &   $3.6\,10^4$ 
               &  $1.9\,10^3$  &   $1.6\,10^3$
                &  $1.3\,10^3$  &   $1.1\,10^3$ & cm$^{-3}$ \\
 $R_{\rm inn}$ (model) &  $8.1\,10^{-3}$  &  $8.1\,10^{-3}$ 
               &  $2.3\,10^{-2}$  &   $2.3\,10^{-2}$ 
                &  $4.9\,10^{-2}$  &  $4.9\,10^{-2}$
                 &  $1.3\,10^{-2}$  &  $1.3\,10^{-2}$ & pc \\
 $\Delta R$ (model)   &  $7.1\,10^{-4}$  &  $2.3\,10^{-2}$ 
               &  $2.8\,10^{-2}$  &   $5.9\,10^{-2}$ 
                &   $9.0\,10^{-2}$  &  $4.1\,10^{-2}$
                 &   $1.9\,10^{-1}$  &  $1.6\,10^{-1}$ & pc \\
 Covering factor &   0.62  &  0.38
                  &  0.17  &  0.83 
                   & 0.0074 &  0.372
                    & 0.028  &  0.437 \\
 $\tau$(13.6\,eV) &  --  & --
                  &  --  &  6.7 
                   &  --  &  1.2
                    &  --  &  3.1 \\
 Mass fraction   &  0.100  & 0.900
                 &  0.108  & 0.892 
                  &  0.092  & 0.907
                   &  0.115  & 0.885 \\
\cline{1-10}        
\noalign{\smallskip}
\end{tabular}       
\end{flushleft}     
\end{table*}        

We used a Cloudy model as an independent check on the NEBU results.  We used
version 05.10.06 of Cloudy, last described by \citet{Ferland1998} and modified
to include additional stellar atmospheres as described below.  The two codes
are independent, and use different assumptions regarding the nebular
structures. A different stellar atmosphere model was used: whereas in the NEBU
code a \citet{Clegg1987} model was adopted, in the Cloudy model we used a He/C
(i.e.  H-poor) model of \citet{Rauch2003}.  Cloudy also allows the use of dust
within the ionized region: we ran models both with and without dust. The
dustless models were marginally better, but in view of the uncertainties in
the modelling, it seems doubtful that this is significant. Cloudy confirmed
the need for a two-component model to obtain a reasonable fit.  In the Cloudy
models the microscopic volume filling factor \citep{Osterbrock1989} was
treated as a free parameter. It converged to unity, tentatively indicating
that the nebular material is homogeneous.  But this required a much less
luminous star than found in the NEBU model.  Radii and density of the
components are also somewhat different.  The oxygen lines are poorly fitted in
comparison to the NEBU results. Convergence to a full solution with Cloudy was
not attempted.  NEBU outputs are presented in Paper~II.

\subsection{The central stars: HR diagram}
\label{hrsection}

The absolute stellar magnitudes, dereddened and converted to Johnson V, are
listed in Table \ref{stellarmag}, for an assumed distance of 25\,kpc.  The
data are corrected for extinction, derived from the H$\alpha$/H$\beta$ ratio.
A standard extinction curve with $R_V=3.1$ is assumed. B-band value are also
given: these are from the long-slit spectroscopy of \citet{Walsh1997} and are
of lower accuracy.

\begin{table}
\caption[]{\label{stellarmag}
Absolute (dereddened) stellar magnitudes for three Sgr PNe, for a distance of
25\,kpc. V-band values are
from the HST data above. B-band values are derived from \citet{Walsh1997}.   
 }
\begin{flushleft}
\begin{tabular}{ l l l l}
\cline{1-4}
\noalign{\smallskip}
  PN         & E(B$-$V) & ~~~~~V  & ~~~B
\\
 \cline{1-4}
\noalign{\smallskip}
He\,2-436      & 0.28 &
         $-0.61\pm0.1$ & $-0.3$: \\
StWr\,2-21     & 0 &
          $+2.70\pm0.1$   \\
Wray\,16-423   & 0.14  &
         $+1.43\pm0.2$  & +1.7:  \\
\cline{1-4}        
\noalign{\smallskip}
\end{tabular}       
\end{flushleft}     
\end{table}

\begin{figure*}
\includegraphics[width=16cm]{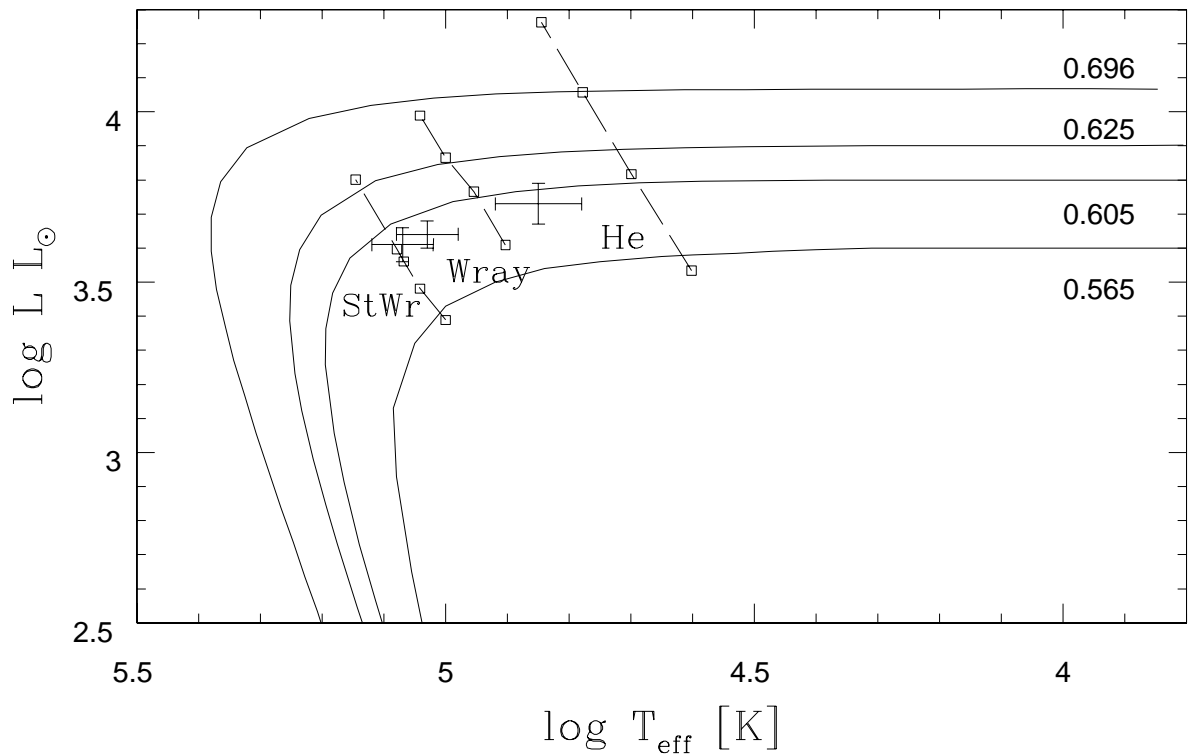}
\caption{\label{hr} Location of the three stars in the HR diagram. The
  solid lines are the theoretical evolutionary models of
  \citet{Bloecker1995b}, labelled with the core mass in solar units.  The long-dashed lines
  show the location of the stars based on the V-band magnitude: from right to
  left He\,2-436, Wray\,16-423 and StWr\,2-21. The square symbols show the
  specific models of \citet{Clegg1987} used in deriving these lines.  The
  photoionization models are shown by the plus signs: in all cases they are
  on or  to the left of the corresponding long-dashed line. }
\end{figure*}

Fig. \ref{hr} shows the location of the stars on the HR diagram, based on the
photoionization models. The uncertainty on the distance is upto 10 per cent,
leading to a systemic uncertainty in log $L$ of 0.1 dex, which is not included
in the error bars. The depth of the Sgr main body has not been measured but is
likely less than 1\,kpc \citep{Majewski2003}.  

The dashed lines in Fig. \ref{hr} show the models of \citet{Bloecker1995b},
labelled by their core mass.  The speed of evolution is a strong function of
core mass. The 'dynamical core mass' in Table \ref{velocityfields} gives the
mass which is consistent with the dynamical age of the nebula. In all three
cases, this mass is around 0.61--0.62 M$_\odot$.  \citep[The systematic
uncertainty in the distance of 10 percent, introduces a $\sim 1$ per cent
error in the core mass, which is much less than the model uncertainties of
$\sim 0.02\,\rm M_\odot$:][]{Gesicki2000}.  The photoionization models yield a
stellar luminosity which is a little lower than corresponds to these masses.
However, the models may underestimate the stellar luminosity if there is
little ionized gas in some directions.

The stellar magnitudes can also be used to derive observed locations in the HR
diagram.  We used the stellar atmosphere models of \citet{Clegg1987} to
calculate predicted V-band magnitudes as function of temperature and
luminosity. The results are shown in Fig. \ref{hr}: the squares show the
luminosities of \citet{Clegg1987} models at specific temperatures, fitted to
the observed V-band magnitudes. The long-dashed lines connect the squares; the
stars are expected to fall along these lines. (In fact, if the dynamical mass
is assumed, the stars should be located on the corresponding intersection
between the long-dashed and the solid lines.)  For He\,2-436, the bright
stellar magnitude indicates a cooler star than given by the photoionization
models.  The most likely explanation is that for this H-poor [WC] star, the
H+He models of \citet{Clegg1987} overpredict the flux at ionizing wavelengths.
For Wray\,16-423 a similar but smaller effect is seen.  The stellar magnitude
of StWr\,2-21 was used as a constraint in the NEBU model.  The relatively
bright star rules out the low luminosity of the Torun model, confirming that
much of the stellar radiation escapes in directions with low density gas as
required in the NEBU model. We note that the results do depend on the
atmosphere model, e.g. using the more modern models of \citet{Rauch2003} could
yield different results. The composition of the stellar atmospheres, and the
existence of a strong wind, introduce a significant additional uncertainty.

\subsection{The central stars: WR lines}
\label{wr_lines}

After subtracting all the nebular emission lines from the spectrum of
StWr\,2-21 by means of spectrum synthesis, two broad features at 3820\AA\ and
5805\AA\ were revealed.  The resulting stellar spectrum is discussed in Paper
II. The presence of broad O\,{\sc vi} 3811+34\AA\ and C\,{\sc iv} 5801+12\AA\ 
emission points to a very hot WR star of subclass [WO] \citep{Acker2003}.  The
intensity ratio in the spectrum synthesis (see paper II) is O\,{\sc vi}
3822/C\,{\sc iv} 5805 $\approx 7.8\pm2$. The uncertainties arise from the
ill-defined 'half line width' of the 3822 line (taken as 1300km/s compared to
750km/s for 5805), and the blending of the stellar 5805 line with (weaker)
nebular 5805 emission.  In a difficult blend, we find C\,{\sc iv} 4658/C\,{\sc
  iv} 5805$ \sim $1--1.5, and there is an indication for O{\, \sc vi}
5290/C\,{\sc iv} 5805 $\sim 0.25$ (where both lines are assumed to have the
same width). These line ratios indicate a [WO\,2] star, perhaps midway between
subclass 2 and 3, according to \citet{Acker2003}.

The two PNe previously known also show WR lines \citep{Walsh1997}.  We
reclassified these stars using the scheme of \citet{Acker2003}, and the data
of \citet{Walsh1997} subject to the reanalysis of \citet{Dudziak2000}.  The
adopted subclasses are listed in Table~\ref{location}. The main constraint is
the absence of some important lines, suggesting they are much weaker than
C\,{\sc iv}. For He\,2-436, the He\,{\sc ii} line points at a possible [WO\,4]
class. Wray\,16-423 is more uncertain, but the detected lines are consistent
with [WC\,4]. The higher temperature of the star of Wray\,16-423 is not
reflected in its subclass. However, this is not unusual among WR nuclei, where
the subclass also traces the strength of the stellar wind
\citep{Crowther1999}. The narrow lines of Wray\,16-423 could also be
interpreted as indicative of a weak emission-line star ({\it wels}).

All three Sgr PN with deep spectra therefore show evidence for WR-type
emission lines.  \citet{Gorny2001} estimates that 6.5 per
cent of Galactic PNe show [WR]-type central stars. A higher fraction is found
in the Galactic Bulge \citep{Gorny2004}, predominantly with late subtypes. To
this should be added the weak emission-line stars \citep[{\it
  wels}:][]{Gesicki2006} which occur  with roughly equal frequency as do
WR-type central stars. To find all three studied
Sgr objects to fit in this category, supports models where the fundamental
parameters of the progenitor star (mass, metallicity) determine whether a
star develops a WR wind. This is in contrast to models where the occurence is
random, caused by the precise timing of the last thermal pulse \citep[for a
review see][]{Bloecker2001}.  There is no evidence for WR-central stars to be
more massive than non-emission-line stars \citep{Pottasch1996}, but
\citet{Gesicki2006} find that Galactic [WC] stars show a narrower,
intermediate range of core masses, compared to non-emission-line stars.  The
Sgr PNe fit within this mass range.

\subsection{The IR-[WC] stars} 

Among Galactic [WC] stars, there is a subgroup with unusually strong IRAS
emission. He\,2-436 has similar characteristics: it is associated with an IRAS
point source \citep{Zijlstra1994}, and scaling its 12-micron flux (0.22\,Jy)
to a typical distance of galactic PNe of 2.5\,kpc would put it among the
brightest such objects. The class of IR-[WC] stars is discussed by
\citet{Zijlstra2001}. 

The Galactic IR-[WC] stars have late subtypes, of 8--11 \citep{Zijlstra2001} .
He\,2-436 shows a much earlier subtype.  Three further extragalactic [WC]
stars are detected by IRAS: SMP58 and SMP61 in the Large Magellanic Cloud and
SMC N6 in the Small Magellanic Cloud \citep{Zijlstra1994, Pena1997,
  Stasinska2004} (the last object has no 12-$\mu$m detection). The two LMC
objects have subclass [WC5-6], and SMP N6 has [WO\,4] on the scale of
\citet{Acker2003}. The extragalactic stars have much earlier subclassses than
comparable Galactic stars. The strong IRAS flux densities require a high dust
optical depth, and indicate a dense, compact nebula. The cause of the
differences between the stars in different galaxies is not clear.

%
%

\subsection{Nebular structures}

\subsubsection{Bipolarity}

The HST images show a very similar global morphology between the three imaged
objects. All show a torus-like structure, with lower-density regions in the
polar directions.  In the classification of \citet{Greig1971}, all three
objects are of class B, indicative of enhanced AGB mass loss in the equatorial
plane. The images do not show the'butterfly' morphology of the more extreme
bipolar nebulae, suggesting the density contrast between equator and poles is
not extreme (in the scheme of \citet{Balick2002}, the objects would be
considered as intermediate between elliptical and bipolar). The fact that all
three nebulae show similar structures is interesting in view of suggestions
that the primary factor determining the nebular morphology is stellar
progenitor mass \citep{Corradi1995}. Reviews on PNe morphologies can be
found in \citet{Kwok2000} and \citet{Balick2002}.

The outer structures of the two larger objects are more complicated;
relatively bright and showing symmetry axes which are not well aligned with
those of the inner torus. The outer regions are sensitive to the mass-loss
evolution on the AGB \citep{Monreal2005} and interaction with the ISM
\citep{Wareing2006}. There is an indication that the asymmetry in the
outer stuctures is along the direction of motion of the galaxy (towards WNW):
interaction with the tenuous medium in the Galactic halo surrounding Sgr
may be affecting the
nebular expansion.

Assuming that the higher-density tori can be associated with the
ionization-bounded sectors in the NEBU models (Table \ref{models}), the tori
of Wray\,16-243 and He\,2-436 contain roughly 10\%\ of the total mass of the
nebulae. This should be taken with care, as the models are imperfect
representations. For StWr\,2-21, the HST image suggests that the torus does
not show a strong ionization front, consistent with the model where this
sector has a very low covering factor.  For Wray\,16-423, the torus has a
clear sharp edge as expected from an ionization front. In both cases, the mass
fraction may be underestimated.  For He\,2-436, the dense sector has a very
small thickness which suggests that the ionization front has barely penetrated
the high density inner shell.  The low total mass of the model also suggests
that a large fraction of its mass is not yet ionized.

The masses in the nebulae are typical compared to
Galactic objects. The bipolar structures are also typical, with half of
Galactic PNe showing similar morphology \citep{Corradi1995}. He\,2-436 has
perhaps a more extreme structure, with a very pronounced, compact torus and a
high density contrast. A number of Galactic post-AGB stars show similar
structures, but these have typically cool and unevolved central stars.

\citet{Stanghellini2003} finds similar statistics on morphology between the
LMC, SMC and the Galaxy.  Our sample agrees with the prevalence of the same
basic morphological types, regardless of environment.  This is an important
result: it provides evidence against models where the (mild) bipolarity
is caused by substellar/planetary companions \citep{Livio2002}, as this would
tend to introduce a metallicity dependence which is not observed.  Strongly
bipolar nebulae may be the product of a unique evolution for each object
\citep{Soker2002}: for the milder stuctures seen here, a single evolutionary
cause may be preferred.

\subsubsection{Double shell}

The presence or absence of a double-shell structure is an important
characteristic of PNe. The evolution of a PN is dominated by a few effects:
(i) the expansion and density profile of the progenitor AGB wind; (ii) the
fast wind from the central star blowing into the AGB wind; (iii) the
ionization of the swept-up nebula and the resultant pressure imbalance 
\citep{Perinotto2004a, Schoenberner2005}. The fast wind sets up a compressed
rim at the inner radius: this is the usual PN. The over-pressure in the
ionized region sets up a swept-up shell  further out. This leads to the
'double shell' consisting of the bright inner rim and the fainter outer
shell. \citet{Perinotto2004a} presents and extensive set of hydrodynamical
models for spherical PNe, to show how this structure evolves as function
of AGB mass loss and central star mass. The double shell appears after the
ionization front breaks out from the inner rim, and the nebula becomes
density-bounded. In earlier phases where the PN is ionization bounded,
only the rim is seen. 

A double shell is evident in the radial profiles of Wray\,16-423 and
StWr\,2-21 (Fig.~\ref{profiles}). The nebula can be approximated by the
hydrodynamical model for a 0.605\,M$_\odot$ central star \citep[sequence 6
of][]{Perinotto2004a}. Here the nebula becomes density bounded after
approximately 3500\,yr. The outer radius of the rim of Wray\,16-423 ($\sim 1.5
\times 10^{17}\,\rm cm$), and the brightness contrast between the rim and
shell of a factor of 4--5, are to first order consistent with the
corresponding Perinotto model in the early density-bounded phase.  (The
spherical asymmetry in the nebulae will cause differences in evolution with
respect to the symetric models.) This model uses a high AGB mass-loss rate,
peaking at $ 3 \times 10^{-4}\rm \, M_\odot\,yr^{-1}$ (case B2 of
\citet{Bloecker1995a} was used). It also has a too high progenitor mass of
3\,M$_\odot$. The size of the rim increases with lower mass-loss rate, but
it is not a strong function. The models suggests that a uniform $\dot M \ge 3
\times 10^{-5}\rm \, M_\odot\,yr^{-1}$ can also be accommodated in order to
explain the observed structure. A high peak mass-loss rate is however more
likely, as AGB mass-loss rates are known to increase with time.

The outer shell of StWr\,2-21 is less distinct because of a low contrast: the
brightness contrast is a factor of 2--3. The Perinotto models show an
increasing contrast with time. As  StWr\,2-21 is the more evolved nebula,
this suggests its evolution has not been identical to that of Wray\,16-423.
However, the differences are relatively minor. The structure of the nebula
changes strongly after the star enters the cooling track, but this has not yet 
happened even for StWr\,2-21 (section \ref{hrsection}).

He\,2-436 shows no indication for such a double shell structure. This is as
expected in its early phase of evolution, where the ionization front is still
trapped inside the rim. The models of \citet{Dudziak2000} require a
central dense inner shell together with a lower-density component. However,
comparison with the images suggests this can be interpreted as the
torus and the polar flows, rather than a radial double-shell structure.

\subsubsection{Dynamics}

The models of \citet{Perinotto2004a} and \citet{Schoenberner2005} also predict
velocity fields, which can be compared to those derived here. The flow
velocities are sensitive to the wind density structure. 

The youngest object, He\,2-436, shows a decreasing velocity with radius over
the highest density gas. As in all cases, the velocities increase again
towards the outer edge (ionization front). The predicted velocities in this
phase show the same behaviour as observed, but the predicted velocities are
around 10\,km\,s$^{-1}$ while the observed expansion is faster. For
Wray\,16-423, a complicated velocity field is found which shows only a vague
resemblance to the predicted one (within the highest density gas, the slight
decline of expansion velocity with radius is consistent with the models, as is
the value of the velocities in this region of around 20\,km\,s$^{-1}$.  The
observed sharp increase near the inner edge is not predicted, nor is the
increase in the low density outer gas.

For StWr\,2-21, the velocities increase linearly with radius, in contrast to
the other two modelled nebulae. This is a predicted feature of more evolved
nebulae, and is a better indication of its evolutionary state than the
density contrast of the rim and shell. The observed velocities (reaching
50\,km\,s$^{-1}$ near the outer edge) are somewhat large compared to the
model values.

The issue of high observed expansion velocities is addressed by
\citet{Schoenberner2005}, who shows that velocities are higher if a steeper
initial density distribution, $\rho \propto r^{-\alpha}$, is assumed.  Our
expansion velocities, compared to Fig. 12 of \citet{Schoenberner2005},
indicate values of $\alpha=2.5$ for He\,2-436, and $\alpha\approx 3$ for the
other two objects. This density distribution is expected from an AGB wind
which increases in time. 

We note that both the Torun models and the Schoenberner models are
one-dimensional. This is a significant limitation for intrinsically 3-d
objects. Observed structure in the velocity field, such as found in
Wray\,16-423, may result from averaging over different directions. The derived
parameters in the Torun models: average expansion velocity, dynamical age and
mass, have been shown to be robust against assumptions regarding nebular
structure, but effect such as turbulence are not. Three-dimensional
kinematic models are being developed \citep[e.g.][]{Morisset2005} but
full hydrodynamical models are still restricted to 1-D.

\subsection{Evolution}

Both stars and nebulae are consistent with an evolutionary sequence where
He-2\,436 is the youngest, Wray\,16-423 more evolved and StWr\,2-21 is the
most evolved nebula, but where the initial conditions were very similar.  The
dynamical ages (Table \ref{velocityfields}) clearly show this sequence; the
nebular radius increases whilst the density drops and the ionization front
moves out. For StWr\,2-21, very little of the mass is still neutral.

The IRAS detection of He\,2-436 \citep{Zijlstra1996} also shows this object to
be young with strong heating of the dust in proximity to the star. The absence
of an IRAS detection for the other two objects is consistent with their much
cooler expected dust.

The comparison with the hydrodynamical models, especially based on
the velocity fields, suggests that the AGB mass-loss rates increased with time,
and reached peak values of $\sim 10^{-4}\,\rm M_\odot \,yr^{-1}$.

\subsection{Abundances}
\label{section_abundances}

Abundances for the Sgr PNe are listed in Table \ref{abundance-comp}.  Values
for He\,2-436 and Wray 16-423 are from \citet{Dudziak2000}. StWr\,2-21
abundances are from the models discussed in Section \ref{photomodel}. The carbon abundance, based on
the C\,{\sc II} recombination line at 4267\AA, is uncertain. The Cloudy model
indicated lower abundances for StWr\,2-21, by about 0.15 dex. We consider the
fully converged NEBU model as more accurate, but either model shows that
StWr\,2-21 has higher abundances than the other PNe.

We also give abundances for BoBn\,1. 
Unlike for the other Sgr PNe, a relatively good IUE spectrum is available for
this PN.  Abundances of BoBn~1 have been obtained by \citet{Howard1997} from
models. However, for  consistency with the other three PNe, the
abundances derived from a two-sector NEBU model worked out by P\'equignot
(unpublished) are quoted in Table~ \ref{abundance-comp}. The N, O and Ne
abundances  are in excellent agreement with a recent compilation by
\citet{Perinotto2004b} (who do not provide S and Ar). The NEBU results on
the other hand differ somewhat (but for different reasons) from either
\citet{Howard1997}  or the re-compilation by the same group \citep{Henry2004}.

Table \ref{abundance-comp} shows the elemental abundances relative to the
solar abundance of \citet{Lodders2003}. These abundances have been
under discussion, and a downward revision to [O/H]\,=\,8.66 has been proposed
\citep{Asplund2004}. This revision has problems reproducing
the known depth of the solar convection zone \citep{Basu2004}.

\begin{table*}
\begin{flushleft}
\caption[]{\label{abundance-comp}
Comparison of abundances, given on a logarithmic scale where H = 12.
}
\begin{tabular}{lrrrrrcccccc}
\hline 
Elem. & \multicolumn{1}{c}{Wray\,16-423} & \multicolumn{1}{c}{He\,2-436} &
 \multicolumn{1}{c}{StWr\,2-21} &   \multicolumn{1}{c}{BoBn\,1} &  
\multicolumn{1}{c}{Galactic} & 
\multicolumn{1}{c}{Solar$^b$$ \odot$} & Wray -- $\odot$ & He~2 -- $\odot$ &
 St -- $\odot$ &  BB -- $\odot$ & GPNe -- $\odot$ \\
 & & & & &
\multicolumn{1}{c}{PNe$^a$} &  &  &  &  \\
\noalign{\smallskip}
\hline
\noalign{\smallskip}
H  & 12.00$\pm$.00 & 12.00 & 12.00  & 12.00$\pm$.00 & 12.00$\pm$.00 &
12.00 &- &- &- &- &- \\ 
He & 11.03$\pm$.01 &  11.03$\pm$.01 & 11.00 & 11.01 & 11.05$\pm$.03 &
10.99 & \  +0.04 & \ +0.04 & \ +0.01 & \ +0.02 & \ +0.06 \\ 
C  &  8.86$\pm$.06 &  9.06$\pm$.09 &  9.00: &  8.83  & 8.81$\pm$.30 & 
8.46  &    +0.40 &   +0.60 & \ +0.54 & \ +0.37 & \ +0.35 \\
N  &  7.68$\pm$.05 &  7.42$\pm$.06 &  7.88 & 7.84   & 8.14$\pm$.20 & 
7.90  & --0.22 &  --0.48 & \ --0.02 & \ --0.06 & \ +0.24 \\
O  &  8.33$\pm$.02 &  8.36$\pm$.06 &  8.53  & 7.75  & 8.69$\pm$.15 & 
8.76  & --0.43 &  --0.40 \ & \ --0.23 & \ --1.01 & \ --0.07 \\
Ne &  7.55$\pm$.03 &  7.54$\pm$.06 &  7.72 & 7.83   & 8.10$\pm$.15 & 
7.95  & --0.40 &  --0.41 \ & \ --0.23 & \ --0.12 & \ +0.15 \\
Mg &  6.98$\pm$.30 &    --         & -- &  --         &             &
7.62       & --0.64: &  --  \ & \ -- & -- &--    \\
S  &  6.67$\pm$.04 &  6.59$\pm$.05 &  6.93 & 5.16    & 6.91$\pm$.30 & 
7.26  & --0.59 &  --0.67 \ & \ --0.33 & \ --2.10 & --0.35 \\
Cl &  4.89$\pm$.18 &      --         & -- &  --         &          &
5.33       & --0.44:  &  -- \ & \ -- &  \ -- & \ --   \\
Ar &  5.95$\pm$.07 &   5.78$\pm$.08 & 6.10 & 4.46       & 6.38$\pm$.30 & 
6.62  & --0.67 &  --0.84  \ & \ --0.52 & \ --2.16 & \ --0.24 \\
K  &  4.65$\pm$.22 &    --         & --  & --         &             &
5.18       & --0.53: &  --  \ & \ -- &  \ -- & \ --  \\
\hline
\end{tabular} 
$^a$ Mean composition and scatter for non-Type I Galactic PNe 
\citep{Kingsburgh1994}\\
$^b$ Solar system abundances are from \citet{Lodders2003}
\end{flushleft}     
\end{table*}

The two previously analyzed PNe have identical abundances, which agree well
with those of the carbon stars of Sgr. StWr\,2-21  has a higher
abundance. BoBn\,1 has a complicated abundance pattern. This is an
extraordinary object in that the neon abundance exceeds that of oxygen.
Such an abundance pattern can only be caused by dredge-up of nuclear-burning
products. The neon enhancement may indicate that the star has suffered a Very
late Thermal Pulse (VLTP), similar to for example Sakurai's Object
\citep{Hajduk2005}. Alternatively, binary coalescence can be considered.
C,N,O and Ne cannot be used to constrain its progenitor abundance: their high
abundances reflect their formation within the star 
\citep{Pequignot2000,Pequignot2005}. 
However, both  the sulfur and argon
abundances of the PN indicate that [Fe/H] $< -2$ and
possibly as low as $-2.3$, considering that, for low
metallicity stars, [Fe/H] tends to be a few 0.1\,dex
smaller than [Ar/H] \citep[e.g.][]{Cayrel2004}.


The [S/O] of the three  PNe in the core of Sgr is $-1.60$ for
StWr\,2-21, and  $-1.70$ and $-1.77$ for the other two objects.
This is close to but perhaps a little lower  than the solar value listed in
Table \ref{abundance-comp} ($-1.59$: the revision of \citet{Asplund2004} would
give an even higher solar ratio).  \citet{Caffau2005} present sulphur
abundances for five stars in Terzan 7, a globular cluster in Sgr with a
metallicity similar to He 2-436 and Wray 16-423. They find [S/O]\,$=-1.71$,
identical to our result. Sulphur and oxygen are both produced in massive
supernovae, but the lower ratio in
the lower metallicity objects may be due to  primary oxygen 
production in low-metallicity AGB stars \citep{Pequignot2000}.

\section{Discussion: The Sgr galaxy}

\subsection{Sgr Luminosity}

We can estimate the luminosity of the Sgr galaxy from its PNe population.  The
specific frequency of PNe is fairly well determined within the Local Group:
\citet{Magrini2003} find $\log N({\rm PN}) = \log L_{\rm V} - 6.92$, in solar
units.  This predicts for Sgr a value of $L_{\rm V} = 10^{7.5}\,\rm L_\odot$,
or $M_{\rm V} = -14.1 $, provided the PNe population in Sgr is now reasonably
complete. The lack of PNe near the centre of Sgr may raise doubt on this,
however the centre is also lacking in globular clusters.  \citet{Majewski2003}
derive $M_{\rm V}=-13.27$ for the main body, assuming a King profile. We
confirm their suspicion that this  underestimates the total
luminosity.

Assuming a mass-to-light ratio of 20--50 \citep{Ibata1997, Majewski2004}, the
total mass of Sgr becomes $M = 1$--$1.5 \times 10^9\,\rm M_\odot$. This is
close to the minimum value required for its survival up to the present
\citep{Ibata1997}.  The fact that one PN is found in the tail makes it likely
that the tails account for a significant fraction of the Sgr mass, but not a
dominant fraction.

\subsection{The enrichment history of Sgr}

StWr\,2-21 shows noticeably high abundances compared to He\,2-436 and
Wray\,16-423. The identical abundances of the latter suggest their progenitor
stars formed in a single, major star-formation burst which took place in a
well-mixed ISM. The burst has been dated at 5\,Gyr ago, based on the AGB
colour, and breaking the age-metallicity degeneracy with the PN abundances
\citep{Dudziak2000}.

Evidence for a  more metal-rich population was presented by
\citet{Bonifacio2004}, based on spectroscopy of 10 red-giant stars.
They derive [Fe/H]$\sim-0.25$. This is in agreement with
StWr\,2-21, for which we find [O/H]$\,=-0.23$. The high metallicity
suggests a recent star formation event. The last major burst
of star formation in Sgr occured between 0.5 and 3 Gyr ago
\citep[e.g.][]{Layden2000}; \citet{Bonifacio2004} date it at  1\,Gyr
or less. (More than one event may have occured.) Assuming 1 and 5\,Gyr,
progenitor masses for the PNe would be 2.2 and 1.3\.M$_\odot$, respectively.

The fact that only one of the PNe shows this high metallicity does not agree
with the finding of \citet{Bonifacio2004} that the large majority of their
stars shows the higher metallicity. Allowance should be made for the small
number of objects in both surveys. The difference could also be related to the
location of their stars: \citet{Bonifacio2004} surveyed a region located
midway between the core region of Sgr and the three PNe.
\citet{Bellazzini2006} find that the large majority of the Sgr population has
a metallicity in the range [M/H]\,=\,$-0.4$ to $-0.7$ and an age of
$8.0\pm1.5$\,Gyr. The more metal-rich, younger population traced by
StWr\,2-21 may not be uniformly distributed.

All four PNe are strongly carbon rich. The morphology of the Sgr dwarf has
been traced extensively with M giants \citep{Majewski2003, Bellazzini2006}.
However, the PNe confirm that even for the most metal-rich and youngest
population, 3rd dredge-up causes the AGB stars to evolve into carbon stars.

BoBn\,1  derives from a much more metal-poor population. The existence
and significance of such a population in Sgr is shown by the metal-poor
clusters M54 (which may be the nucleus of Sgr) and Arp2 have metallicity
[Fe/H]\,=\,$-1.8$ \citep{Layden2000}, and Ter 8 is reported as
[Fe/H]\,=\,$-2.0$ \citep{Layden1997}. The RR Lyrae population also shows 
a minor but significant population with [Fe/H]$\lsim-2$
\citep{Cseresnjes2001}. BoBn\,1, if its association with Sgr is confirmed,
may derive from this earliest population of Sgr.

\subsection{Evolution of Sgr}

The abundance distributions of Dwarf Spheroidal galaxies, including Sgr,
differ from dwarf irregular galaxies of similar luminosities.  Both contain
old stars of very low abundances \citep[which in dwarf irregulars show a more
spheroidal-like distribution:][]{Minniti1996}. But whereas younger stars
in dIrr galaxies still are at low abundances, with a strong relation between
[O/H] and galaxy luminosity \citep{Pilyugin2001}, for dSph galaxies younger
stars are more metal-rich \citep{Grebel2003}. For galaxies of similar mass and
luminosities, the dSph galaxies have a dominant population which is much more
metal-rich than do dIrr galaxies \citep[e.g.][]{Mateo1998}. Different reasons
have been suggested: more efficient enrichment, less infall of unprocessed gas
or a large amount of tidal stripping \citep{Grebel2003}.  dIrr galaxies may
have low efficiency in enrichment due to blow-out of stellar ejecta by
supernovae, or they may benefit from continuous infall of unprocessed
extra-galactic HI. dSph galaxies may have lacked any subsequent gas infall
(\citet{Layden2000} argue for a closed-box chemical evolution in Sgr). Present
dSph systems may also have had a much higher luminosity/mass in the past with
the present galaxy being only a remnant of the original system. It is however
likely that the cause is related to the location of the galaxies: dwarf
spheroidals are found near large galaxies while dwarf irregulars are more
isolated.
 
The ISM in which the recent star formation of Sgr occured is unlikely to have
been original. The orbit of Sgr takes it through the Galactic plane, outside
the solar circle, every 1Gyr or so. This is believed to be a possible cause of
sudden star formation bursts, but the passage will also strip Sgr of its gas.
At present there is no evidence for ISM within the main body of Sgr although
stripped HI has been found along its orbit \citep{Putman2004}.

Two sources of ISM renewal can be considered. The first is mass loss from
stars, as traced by the planetary nebulae. The three PNe in the main body of
Sgr suggest a current stellar death rate of about $10^{-3}\,\rm yr^{-1}$.  For
progenitor masses of 1.3--1.5 M$_\odot$, and stellar remnants of 0.6
M$_\odot$, the mass return rate to the ISM becomes $5\times 10^{-4}\,M_\odot\,
\rm yr^{-1} $. Over 1\,Gyr, the approximate orbital period of Sgr, the ISM can
reach $\sim 10^6\,\rm M_\odot$ which may be sufficient to trigger star
formation.  In the absence of significant rotation in the stellar population,
this lost mass will congregate in the potential well. Thus, the next star
formation event will occur near the centre of the galaxy, as is observed in
Fornax \citep{Saviane2000}. As the original low metallicity ISM has been
preferentially removed, the enrichment efficiency will be considerably higher
than in the closed-box model. All these points are consistent with the
properties of the dSph galaxies, including Sgr.

A second possiblity to be considered is that the dSph captured gas from
the Galactic halo. Assuming that the halo contains a reservoir of
cooling gas originating from a Galactic Fountain, a galaxy in orbit
within the halo can be expected to capture some of this gas. This
can explain the very high metallicities, as the halo gas
has metallicities appropriate for the host galaxy. In combination with
the self-enrichment scenario above, this would give an even faster
increase of metallicity with time.  It is noticeable that the
metallicity of the most recent star formation event in Sgr has a
metallicity, as shown by StWr\,2-21, not below that of regions of the
outer Galactic plane.

We finally note the similarity with the Fornax dwarf, a galaxy with similar
mass and properties to Sgr. It also shows evidence for an extended star
formation history with increasing metallicity. Both the old and the youngest
population (at $\sim 2\,10^8$\,yr even younger than Sgr) have similar
metallicity to their counterparts in Sgr, [Fe/H]\,$=-0.4$, to $-2.0$
\citep{Saviane2000}. Fornax has suffered much less ram pressure stripping and
tidal stripping from the Galaxy, being in a much more distant orbit.  This
suggests that the fast chemical evolution is not dominated by gas removal
through ram pressure.  Instead gas accretion, or lack thereof, may drive the
evolution. Enrichment in Fornax and Sgr may have been accelerated by a lack of
an external source of metal-poor gas (as compared to isolated dIrr galaxies),
or by accretion of enriched gas expelled by our Galaxy.

The origin of the difference in enrichment history between dSph and dIrr
galaxies is still an enigma \citep{Grebel2003}.  It is worth testing whether
the isolated transition galaxies between dIrr and dSph, such as Phoenix,
happen to be located in regions of a low density of accretable, non-enriched
gas.

\subsection{The tidal tails}

The tidal tails of Sgr  encircle the Milky Way Galaxy in a complicated pattern.
The system clearly has had a long history of disruption. The models
of \citet{Law2005} show how different parts of the tails  trace different
distortion events.

The Sgr tidal tails are an important, but not dominant, contribution to the
Galactic halo. On statistical grounds, \citet{Bellazzini2003} argue that 4 of
the 35 halo globular clusters should be linked with Sgr debris. Since
approximately 10 halo PNe are known, our finding that one of these (BoBn\,1)
is a tidal member of Sgr is in line with the globular cluster fraction.

\citet{Bellazzini2003} do not consider the globular clusters beyond 60\,kpc,
as falling beyond the apogee of the Sgr orbit. But the tidal streams are found
out to 80--100\,kpc \citep{Law2005} and at least one of their discarded
distant clusters (NGC\,2419) is attributed to Sgr debris \citep{Newberg2003}.
No halo PNe are known at anywhere near such distances, possibly because of
faintness.

The tidal tails trace the older, metal-poor population of Sgr
\citep{Alard2001, Bellazzini2002}, as shown by its RR Lyrae stars
\citep{Cseresnjes2001}. (\citet{Bellazzini2006} argue that the dominant
metal-rich population should also be distributed around the tidal tails.)  The
globular clusters indicate that the [Fe/H] of the tidal tails is around $-1.5$
to $-1.9$. This population becomes dominant in the tails because of the
age-metallicity gradient: the earlier disruptions affected the older, less
metal-rich population.

Within the Local Group, populations with [Fe/H]\,$\lsim -1$ show indications
for a deficit of PNe \citep{Zijlstra2004, Magrini2003}, related to a lower
mass-loss efficiency on the AGB. This would enhance the fraction of peculiar
objects (e.g. VLTP nebulae, for which the formation rate does not depend on
metallicity), to which category BoBn\,1 likely belongs.

\section{Conclusions}

The main points of our study of the PNe population of the Sagittarius dwarf
spheroidal galaxy can be summarized as follows.

1. Two new PNe are discovered in Sgr, bringing the total number to four. One
is located in the leading tidal tail, the other three are located in the main
body, south of the main centre of Sgr.  The PNe population (assuming the
sample is complete) indicates an absolute visual magnitude of $M_{\rm
  V}=-14.1$ for Sgr, including its leading and trailing tails. A stellar death
rate of $10^{-3}$ per year is derived.

2. HST images of the three PNe in the main body of Sgr show very similar,
bipolar morphologies, consisting of a denser equatorial torus and lower
density, more extended polar regions. This type of morphology is also common
among Galactic PNe, and therefore the origin is unlikely to be dependent on
metallicity-dependent effects, as would be expected if planetary companions
provide the angular momentum required for the equatorially-enhanced mass loss.
Whereas for strongly bipolar PNe it has been argued that every object is the
product of a unique evolution \citep{Soker2002}, for the mildly bipolar
nebulae a single cause is preferred.

3. Analysis of the central stars shows that all are on the post-AGB track,
 prior to entering the cooling track. The effective temperatures are between
  $7 \times 10^4$ and $1.1 \times 10^5$\,K, and masses are close to
 0.61\,M$_\odot$.  Post-AGB ages are between 1000 and 5000 years.
 
 4. All three analyzed PNe show WR emission lines, with early subclasses of
 [WO\,2] to [WC\,4]. WR-type central stars of PNe form a minority population
 among Galactic PNe. Their prominence in the Sgr galaxy suggests that the
 progenitor star characteristics (mass, metallicity) determine the occurence
 of such stars. This is in contrast to models where the (random) timing of the
 last thermal pulse is the crucial factor.
 
 5. Photoionization models for the three PNe with HST images yield abundances
 of [O/H]$\,=-0.23$ for one object and [O/H]$\,=-0.55$ for the other two. The
 first object confirms the reality of the metal-rich population uncovered
 recently by \citet{Bonifacio2004}, and suggests it accounts for a substantial
 part of the Sgr stellar population, albeit not its majority.  The identical
 abundances of the other two objects argue for a short-lived burst in a
 well-mixed ISM: this burst has been dated at 5\,Gyr ago. The higher
 metallicity object may date from a burst roughly 1\,Gyr ago. The fourth
 object, BoBn\,1, located in the tidal tail, has very unusual abundances and a
 non-standard history, but a progenitor metallicity of [Fe/H]$\sim-2$ is
 indicated. This object may belong to the oldest population of Sgr.

6.  The mass-weighted expansion 
velocities are similar to those of Galactic PNe, and show no indication of
a metallicity effect. Kinematic ages and core masses are derived: all three
objects have similar core masses of around 0.61\,M$_\odot$.  The age--velocity
relation, compared to hydrodynamical models, suggest a steep initial density
distribution, of $\rho \propto r^{-3}$. This indicates that the preceding AGB
stars showed mass-loss rates increasing with time.
 
7. The ISM reformation rate of Sgr due to stellar mass loss is $5 \times
10^{-4}\, \rm M_\odot\, yr^{-1}$. This amounts to $10^6\,\rm M_\odot$
over one orbit, which may be sufficient to trigger renewed star formation.

8. The steep metallicity-age gradient, and the very high recent abundance,
must be due to a combination of stripping of older ISM from Sgr, with
enrichment from its own stellar mass loss. The fact that Fornax, which is in a
much wider orbit around the Milky Way, shows very similar recent enrichment,
may indicate that ram-pressure stripping is not the dominant effect. We
suggest that accretion of gas originating from the Milky Way may play a role.
This is based on the fact that the most metal-rich population has a
metallicity not too different from that of the outer regions of the Milky Way
Galaxy.

 9. We finally note that He\,2-436 provides the sole direct detection of dust
 in a dwarf spheroidal galaxy, to date.

\section*{Acknowledgements}
This research was supported by a NATO collaboratve program grant number
PST.CLG.979726. KG acknowledges financial support from the Polish State
Committee for Scientific Research grant No. 2.P.03D.002.025. PvH acknowledges
support from the Belgian Science Policy Office under IAP grant P5/36.  DM is
partially supported by FONDAP 15010003 and by the John Simon Gugenheim
Foundation. ESO supported this research through its visitor program. We
gratefully acknowledge discussion with C. Morisset. The HST data was part of
observing program 9356. 

\bibliographystyle{mn2e}

\bibliography{paper-sgr}

\begin{thebibliography}{}

\bibitem[\protect\citeauthoryear{{Acker}, {Marcout}, {Ochsenbein}, {Stenholm}
  \& {Tylenda}}{{Acker} et~al.}{1992}]{Acker1992}
{Acker} A.,  {Marcout} J.,  {Ochsenbein} F.,  {Stenholm} B.,    {Tylenda} R.,
  1992, {Strasbourg - ESO catalogue of galactic planetary nebulae.}.
Garching: European Southern Observatory

\bibitem[\protect\citeauthoryear{{Acker} \& {Neiner}}{{Acker} \&
  {Neiner}}{2003}]{Acker2003}
{Acker} A.,  {Neiner} C.,  2003, A\&A, 403, 659

\bibitem[\protect\citeauthoryear{{Alard}}{{Alard}}{1996}]{Alard1996}
{Alard} C.,  1996, ApJ, 458, L17

\bibitem[\protect\citeauthoryear{{Alard}}{{Alard}}{2001}]{Alard2001}
{Alard} C.,  2001, A\&A, 377, 389

\bibitem[\protect\citeauthoryear{{Asplund}, {Grevesse}, {Sauval}, {Allende
  Prieto} \& {Kiselman}}{{Asplund} et~al.}{2004}]{Asplund2004}
{Asplund} M.,  {Grevesse} N.,  {Sauval} A.~J.,  {Allende Prieto} C.,
  {Kiselman} D.,  2004, A\&A, 417, 751

\bibitem[\protect\citeauthoryear{{Balick} \& {Frank}}{{Balick} \&
  {Frank}}{2002}]{Balick2002}
{Balick} B.,  {Frank} A.,  2002, ARA\&A, 40, 439

\bibitem[\protect\citeauthoryear{{Basu} \& {Antia}}{{Basu} \&
  {Antia}}{2004}]{Basu2004}
{Basu} S.,  {Antia} H.~M.,  2004, ApJ, 606, L85

\bibitem[\protect\citeauthoryear{{Bellazzini}, {Correnti}, {Ferraro}, {Monaco}
  \& P.}{{Bellazzini} et~al.}{2006}]{Bellazzini2006}
{Bellazzini} M.,  {Correnti} M.,  {Ferraro} F.~R.,  {Monaco} L.,    P. M.,
  2006, A\&A, 446, L1

\bibitem[\protect\citeauthoryear{{Bellazzini}, {Ferraro} \&
  {Ibata}}{{Bellazzini} et~al.}{2002}]{Bellazzini2002}
{Bellazzini} M.,  {Ferraro} F.~R.,    {Ibata} R.,  2002, AJ, 124, 915

\bibitem[\protect\citeauthoryear{{Bellazzini}, {Ferraro} \&
  {Ibata}}{{Bellazzini} et~al.}{2003}]{Bellazzini2003}
{Bellazzini} M.,  {Ferraro} F.~R.,    {Ibata} R.,  2003, AJ, 125, 188

\bibitem[\protect\citeauthoryear{{Bl{oe}cker}}{{Bl\"ocker}}{2001}]{Bloecker2001}
{Bl\"ocker} T.,  2001, Ap\&SS, 275, 1

\bibitem[\protect\citeauthoryear{{Bl\"ocker}}{{Bl\"ocker}}{1995a}]{Bloecker1995a}
{Bl\"ocker} T.,  1995a, A\&A, 297, 727

\bibitem[\protect\citeauthoryear{{Bloecker}}{{Bl\"ocker}}{1995b}]{Bloecker1995b}
{Bl\"ocker} T.,  1995b, A\&A, 299, 755

\bibitem[\protect\citeauthoryear{{Bonifacio}, {Sbordone}, {Marconi}, {Pasquini}
  \& {Hill}}{{Bonifacio} et~al.}{2004}]{Bonifacio2004}
{Bonifacio} P.,  {Sbordone} L.,  {Marconi} G.,  {Pasquini} L.,    {Hill} V.,
  2004, A\&A, 414, 503

\bibitem[\protect\citeauthoryear{{Caffau}, {Bonifacio}, {Faraggiana} \&
  {Sbordone}}{{Caffau} et~al.}{2005}]{Caffau2005}
{Caffau} E.,  {Bonifacio} P.,  {Faraggiana} R.,    {Sbordone} L.,  2005, A\&A,
  436, L9

\bibitem[\protect\citeauthoryear{{Cayrel}, {Depagne}, {Spite}, {Hill}, {Spite},
  {Fran{\c c}ois}, {Plez}, {Beers}, {Primas}, {Andersen}, {Barbuy},
  {Bonifacio}, {Molaro} \& {Nordstr{\"o}m}}{{Cayrel} et~al.}{2004}]{Cayrel2004}
{Cayrel} R.,  {Depagne} E.,  {Spite} M.,  {Hill} V.,  {Spite} F.,  {Fran{\c
  c}ois} P.,  {Plez} B.,  {Beers} T.,  {Primas} F.,  {Andersen} J.,  {Barbuy}
  B.,  {Bonifacio} P.,  {Molaro} P.,    {Nordstr{\"o}m} B.,  2004, A\&A, 416,
  1117

\bibitem[\protect\citeauthoryear{{Clegg} \& {Middlemass}}{{Clegg} \&
  {Middlemass}}{1987}]{Clegg1987}
{Clegg} R.~E.~S.,  {Middlemass} D.,  1987, MNRAS, 228, 759

\bibitem[\protect\citeauthoryear{{Corradi} \& {Schwarz}}{{Corradi} \&
  {Schwarz}}{1995}]{Corradi1995}
{Corradi} R.~L.~M.,  {Schwarz} H.~E.,  1995, A\&A, 293, 871

\bibitem[\protect\citeauthoryear{{Crowther}}{{Crowther}}{1999}]{Crowther1999}
{Crowther} P.~A.,  1999, in IAU Symp. 193: Wolf-Rayet Phenomena in Massive
  Stars and Starburst Galaxies, p.~116

\bibitem[\protect\citeauthoryear{{Cseresnjes}}{{Cseresnjes}}{2001}]{Cseresnjes%
2001}
{Cseresnjes} P.,  2001, A\&A, 375, 909

\bibitem[\protect\citeauthoryear{{Dohm-Palmer}, {Helmi}, {Morrison}, {Mateo},
  {Olszewski}, {Harding}, {Freeman}, {Norris} \& {Shectman}}{{Dohm-Palmer}
  et~al.}{2001}]{DohmPalmer2001}
{Dohm-Palmer} R.~C.,  {Helmi} A.,  {Morrison} H.,  {Mateo} M.,  {Olszewski}
  E.~W.,  {Harding} P.,  {Freeman} K.~C.,  {Norris} J.,    {Shectman} S.~A.,
  2001, ApJ, 555, L37

\bibitem[\protect\citeauthoryear{{Dudziak}, {P{\' e}quignot}, {Zijlstra} \&
  {Walsh}}{{Dudziak} et~al.}{2000}]{Dudziak2000}
{Dudziak} G.,  {P{\' e}quignot} D.,  {Zijlstra} A.~A.,    {Walsh} J.~R.,  2000,
  A\&A, 363, 717

\bibitem[\protect\citeauthoryear{{Ferland}, {Korista}, {Verner}, {Ferguson},
  {Kingdon} \& {Verner}}{{Ferland} et~al.}{1998}]{Ferland1998}
{Ferland} G.~J.,  {Korista} K.~T.,  {Verner} D.~A.,  {Ferguson} J.~W.,
  {Kingdon} J.~B.,    {Verner} E.~M.,  1998, PASP, 110, 761

\bibitem[\protect\citeauthoryear{{G{\' o}rny}}{{G{\' o}rny}}{2001}]{Gorny2001}
{G{\' o}rny} S.~K.,  2001, Ap\&SS, 275, 67

\bibitem[\protect\citeauthoryear{{G{\' o}rny}, {Stasi{\' n}ska}, {Escudero} \&
  {Costa}}{{G{\' o}rny} et~al.}{2004}]{Gorny2004}
{G{\' o}rny} S.~K.,  {Stasi{\' n}ska} G.,  {Escudero} A.~V.,    {Costa}
  R.~D.~D.,  2004, A\&A, 427, 231

\bibitem[\protect\citeauthoryear{{Gesicki} \& {Zijlstra}}{{Gesicki} \&
  {Zijlstra}}{2000}]{Gesicki2000}
{Gesicki} K.,  {Zijlstra} A.~A.,  2000, A\&A, 358, 1058

\bibitem[\protect\citeauthoryear{{Gesicki} \& {Zijlstra}}{{Gesicki} \&
  {Zijlstra}}{2003}]{Gesicki2003a}
{Gesicki} K.,  {Zijlstra} A.~A.,  2003, MNRAS, 338, 347

\bibitem[\protect\citeauthoryear{{Gesicki}, {Zijlstra}, {Acker}, {G\'orny},
  {Gozdziewski} \& {Walsh}}{{Gesicki} et~al.}{2006}]{Gesicki2006}
{Gesicki} K.,  {Zijlstra} A.~A.,  {Acker} A.,  {G\'orny} S.~K.,  {Gozdziewski}
  K.,    {Walsh} J.~R.,  2006, A\&A, in press

\bibitem[\protect\citeauthoryear{{Grebel}, {Gallagher} \& {Harbeck}}{{Grebel}
  et~al.}{2003}]{Grebel2003}
{Grebel} E.~K.,  {Gallagher} J.~S.,    {Harbeck} D.,  2003, AJ, 125, 1926

\bibitem[\protect\citeauthoryear{{Greig}}{{Greig}}{1971}]{Greig1971}
{Greig} W.~E.,  1971, A\&A, 10, 161

\bibitem[\protect\citeauthoryear{{Hajduk}, {Zijlstra}, {Herwig}, {van Hoof},
  {Kerber}, {Kimeswenger}, {Pollacco}, {Evans}, {Lop{\' e}z}, {Bryce}, {Eyres}
  \& {Matsuura}}{{Hajduk} et~al.}{2005}]{Hajduk2005}
{Hajduk} M.,  {Zijlstra} A.~A.,  {Herwig} F.,  {van Hoof} P.~A.~M.,  {Kerber}
  F.,  {Kimeswenger} S.,  {Pollacco} D.~L.,  {Evans} A.,  {Lop{\' e}z} J.~A.,
  {Bryce} M.,  {Eyres} S.~P.~S.,    {Matsuura} M.,  2005, Science, 308, 231

\bibitem[\protect\citeauthoryear{{Hawley} \& {Miller}}{{Hawley} \&
  {Miller}}{1978}]{Hawley1978}
{Hawley} S.~A.,  {Miller} J.~S.,  1978, ApJ, 220, 609

\bibitem[\protect\citeauthoryear{{Helmi} \& {White}}{{Helmi} \&
  {White}}{2001}]{Helmi2001}
{Helmi} A.,  {White} S.~D.~M.,  2001, MNRAS, 323, 529

\bibitem[\protect\citeauthoryear{{Henry}, {Kwitter} \& {Balick}}{{Henry}
  et~al.}{2004}]{Henry2004}
{Henry} R.~B.~C.,  {Kwitter} K.~B.,    {Balick} B.,  2004, AJ, 127, 2284

\bibitem[\protect\citeauthoryear{{Howard}, {Henry} \& {McCartney}}{{Howard}
  et~al.}{1997}]{Howard1997}
{Howard} J.~W.,  {Henry} R.~B.~C.,    {McCartney} S.,  1997, MNRAS, 284, 465

\bibitem[\protect\citeauthoryear{{Ibata}, {Lewis}, {Irwin}, {Totten} \&
  {Quinn}}{{Ibata} et~al.}{2001}]{Ibata2001}
{Ibata} R.,  {Lewis} G.~F.,  {Irwin} M.,  {Totten} E.,    {Quinn} T.,  2001,
  ApJ, 551, 294

\bibitem[\protect\citeauthoryear{{Ibata}, {Gilmore} \& {Irwin}}{{Ibata}
  et~al.}{1994}]{Ibata1994}
{Ibata} R.~A.,  {Gilmore} G.,    {Irwin} M.~J.,  1994, Nature, 370, 194

\bibitem[\protect\citeauthoryear{{Ibata}, {Wyse}, {Gilmore}, {Irwin} \&
  {Suntzeff}}{{Ibata} et~al.}{1997}]{Ibata1997}
{Ibata} R.~A.,  {Wyse} R.~F.~G.,  {Gilmore} G.,  {Irwin} M.~J.,    {Suntzeff}
  N.~B.,  1997, AJ, 113, 634

\bibitem[\protect\citeauthoryear{{Jiang} \& {Binney}}{{Jiang} \&
  {Binney}}{2000}]{Jiang2000}
{Jiang} I.,  {Binney} J.,  2000, MNRAS, 314, 468

\bibitem[\protect\citeauthoryear{{Kingsburgh} \& {Barlow}}{{Kingsburgh} \&
  {Barlow}}{1994}]{Kingsburgh1994}
{Kingsburgh} R.~L.,  {Barlow} M.~J.,  1994, MNRAS, 271, 257

\bibitem[\protect\citeauthoryear{{Kwok}}{{Kwok}}{2000}]{Kwok2000}
{Kwok} S., 2000, {The origin and evolution of planetary nebulae},
{Cambridge University Press}

\bibitem[\protect\citeauthoryear{{Law}, {Johnston} \& {Majewski}}{{Law}
  et~al.}{2005}]{Law2005}
{Law} D.~R.,  {Johnston} K.~V.,    {Majewski} S.~R.,  2005, ApJ, 619, 807

\bibitem[\protect\citeauthoryear{{Layden} \& {Sarajedini}}{{Layden} \&
  {Sarajedini}}{1997}]{Layden1997}
{Layden} A.~C.,  {Sarajedini} A.,  1997, ApJ, 486, L107

\bibitem[\protect\citeauthoryear{{Layden} \& {Sarajedini}}{{Layden} \&
  {Sarajedini}}{2000}]{Layden2000}
{Layden} A.~C.,  {Sarajedini} A.,  2000, AJ, 119, 1760

\bibitem[\protect\citeauthoryear{{Liu}, {Liu}, {Barlow} \& {Luo}}{{Liu}
  et~al.}{2004}]{Liu2004}
{Liu} Y.,  {Liu} X.-W.,  {Barlow} M.~J.,    {Luo} S.-G.,  2004, MNRAS, 353,
  1251

\bibitem[\protect\citeauthoryear{{Livio} \& {Soker}}{{Livio} \&
  {Soker}}{2002}]{Livio2002}
{Livio} M.,  {Soker} N.,  2002, ApJ, 571, L161

\bibitem[\protect\citeauthoryear{{Lodders}}{{Lodders}}{2003}]{Lodders2003}
{Lodders} K.,  2003, ApJ, 591, 1220

\bibitem[\protect\citeauthoryear{{Magrini}, {Corradi}, {Greimel}, {Leisy},
  {Lennon}, {Mampaso}, {Perinotto}, {Pollacco}, {Walsh}, {Walton} \&
  {Zijlstra}}{{Magrini} et~al.}{2003}]{Magrini2003}
{Magrini} L.,  {Corradi} R.~L.~M.,  {Greimel} R.,  {Leisy} P.,  {Lennon} D.~J.,
   {Mampaso} A.,  {Perinotto} M.,  {Pollacco} D.~L.,  {Walsh} J.~R.,  {Walton}
  N.~A.,    {Zijlstra} A.~A.,  2003, A\&A, 407, 51

\bibitem[\protect\citeauthoryear{{Majewski}, {Kunkel}, {Law}, {Patterson},
  {Polak}, {Rocha-Pinto}, {Crane}, {Frinchaboy}, {Hummels}, {Johnston}, {Rhee},
  {Skrutskie} \& {Weinberg}}{{Majewski} et~al.}{2004}]{Majewski2004}
{Majewski} S.~R.,  {Kunkel} W.~E.,  {Law} D.~R.,  {Patterson} R.~J.,  {Polak}
  A.~A.,  {Rocha-Pinto} H.~J.,  {Crane} J.~D.,  {Frinchaboy} P.~M.,  {Hummels}
  C.~B.,  {Johnston} K.~V.,  {Rhee} J.,  {Skrutskie} M.~F.,    {Weinberg} M.,
  2004, AJ, 128, 245

\bibitem[\protect\citeauthoryear{{Majewski}, {Skrutskie}, {Weinberg} \&
  {Ostheimer}}{{Majewski} et~al.}{2003}]{Majewski2003}
{Majewski} S.~R.,  {Skrutskie} M.~F.,  {Weinberg} M.~D.,    {Ostheimer} J.~C.,
  2003, ApJ, 599, 1082

\bibitem[\protect\citeauthoryear{{Marshall}, {van Loon}, {Matsuura}, {Wood},
  {Zijlstra} \& {Whitelock}}{{Marshall} et~al.}{2004}]{Marshall2004}
{Marshall} J.~R.,  {van Loon} J.~T.,  {Matsuura} M.,  {Wood} P.~R.,  {Zijlstra}
  A.~A.,    {Whitelock} P.~A.,  2004, MNRAS, 355, 1348

\bibitem[\protect\citeauthoryear{{Mateo}}{{Mateo}}{1998}]{Mateo1998}
{Mateo} M.~L.,  1998, ARA\&A, 36, 435

\bibitem[\protect\citeauthoryear{{Mauron}, {Kendall} \& {Gigoyan}}{{Mauron}
  et~al.}{2005}]{Mauron2005}
{Mauron} N.,  {Kendall} T.~R.,    {Gigoyan} K.,  2005, A\&A, 438, 867

\bibitem[\protect\citeauthoryear{{Minniti} \& {Zijlstra}}{{Minniti} \&
  {Zijlstra}}{1996}]{Minniti1996}
{Minniti} D.,  {Zijlstra} A.~A.,  1996, ApJ, 467, L13

\bibitem[\protect\citeauthoryear{{Monaco}, {Bellazzini}, {Ferraro} \&
  {Pancino}}{{Monaco} et~al.}{2004}]{Monaco2004}
{Monaco} L.,  {Bellazzini} M.,  {Ferraro} F.~R.,    {Pancino} E.,  2004, MNRAS,
  353, 874

\bibitem[\protect\citeauthoryear{{Monreal-Ibero}, {Roth}, {Sch{\" o}nberner},
  {Steffen} \& {B{\" o}hm}}{{Monreal-Ibero} et~al.}{2005}]{Monreal2005}
{Monreal-Ibero} A.,  {Roth} M.~M.,  {Sch{\" o}nberner} D.,  {Steffen} M.,
  {B{\" o}hm} P.,  2005, ApJ, 628, L139

\bibitem[\protect\citeauthoryear{{Morisset}, {Stasinska} \& {Pena}}{{Morisset}
  et~al.}{2005}]{Morisset2005}
{Morisset} C.,  {Stasinska} G.,    {Pena} M.,  2005, in Szczerba R.,  Stasinska
  G.,   Gorny S.,  eds, Planetary Nebulae as Astronomical Tools, AIP Conf.
  Proc., Vol.~804.
p.~44

\bibitem[\protect\citeauthoryear{{Newberg}, {Yanny}, {Grebel}, {Hennessy},
  {Ivezi{\' c}}, {Martinez-Delgado}, {Odenkirchen}, {Rix}, {Brinkmann}, {Lamb},
  {Schneider} \& {York}}{{Newberg} et~al.}{2003}]{Newberg2003}
{Newberg} H.~J.,  {Yanny} B.,  {Grebel} E.~K.,  {Hennessy} G.,  {Ivezi{\' c}}
  {\v Z}.,  {Martinez-Delgado} D.,  {Odenkirchen} M.,  {Rix} H.-W.,
  {Brinkmann} J.,  {Lamb} D.~Q.,  {Schneider} D.~P.,    {York} D.~G.,  2003,
  ApJ, 596, L191

\bibitem[\protect\citeauthoryear{{O'Dell} \& {Doi}}{{O'Dell} \&
  {Doi}}{1999}]{ODell1999}
{O'Dell} C.~R.,  {Doi} T.,  1999, PASP, 111, 1316

\bibitem[\protect\citeauthoryear{{Oke}}{{Oke}}{1990}]{Oke1990}
{Oke} J.~B.,  1990, AJ, 99, 1621

\bibitem[\protect\citeauthoryear{{Osterbrock}}{{Osterbrock}}{1989}]{Osterbrock%
1989}
{Osterbrock} D.~E.,  1989, {Astrophysics of gaseous nebulae and active galactic
  nuclei}.
University of California, University Science Books

\bibitem[\protect\citeauthoryear{{Parker}, {Hartley}, {Russeil}, {Acker},
  {Morgan}, {Beaulieu}, {Morris}, {Phillips} \& {Cohen}}{{Parker}
  et~al.}{2003}]{Parker2003}
{Parker} Q.~A.,  {Hartley} M.,  {Russeil} D.,  {Acker} A.,  {Morgan} D.~H.,
  {Beaulieu} S.,  {Morris} R.,  {Phillips} S.,    {Cohen} M.,  2003, in S.~Kwok
  M.~D.,  Sutherland R.,  eds, Planetary Nebulae: Their Evolution and Role in
  the Universe, IAU Symposium, Vol.~209, {}.
Astron. Soc. Pac., p.~25

\bibitem[\protect\citeauthoryear{{Patat}}{{Patat}}{1990}]{Patat1990}
{Patat} F.,  1990, {EFOSC2 User's Manual}.
ESO, Garching

\bibitem[\protect\citeauthoryear{{Pena}, {Ruiz} \& {Torres-Peimbert}}{{Pena}
  et~al.}{1997}]{Pena1997}
{Pena} M.,  {Ruiz} M.~T.,    {Torres-Peimbert} S.,  1997, A\&A, 324, 674

\bibitem[\protect\citeauthoryear{{P{\'e}quignot}, {Liu}, {Barlow}, {Storey} \&
  {Morisset}}{{P{\'e}quignot} et~al.}{2003}]{Pequignot2003}
{P{\'e}quignot} D.,  {Liu} X.-W.,  {Barlow} M.~J.,  {Storey} P.~J.,
  {Morisset} C.,  2003, in IAU Symposium 209: Planetary Nebulae: Their
  Evolution and Role in the Universe, Eds.  Sun Kwok, Michael Dopita, and Ralph Sutherland.
p.~347

\bibitem[\protect\citeauthoryear{{P{\' e}quignot}, {Walsh}, {Zijlstra} \&
  {Dudziak}}{{P{\' e}quignot} et~al.}{2000}]{Pequignot2000}
{P{\' e}quignot} D.,  {Walsh} J.~R.,  {Zijlstra} A.~A.,    {Dudziak} G.,  2000,
  A\&A, 361, L1

\bibitem[\protect\citeauthoryear{{P{\'e}quignot} \& {Tsamis}}{{P{\'e}quignot}
  \& {Tsamis}}{2005}]{Pequignot2005}
{P{\'e}quignot} D.,  {Tsamis} Y.~G.,  2005, A\&A, 430, 187

\bibitem[\protect\citeauthoryear{{Perinotto}, {Morbidelli} \&
  {Scatarzi}}{{Perinotto} et~al.}{2004}]{Perinotto2004b}
{Perinotto} M.,  {Morbidelli} L.,    {Scatarzi} A.,  2004, MNRAS, 349, 793

\bibitem[\protect\citeauthoryear{{Perinotto}, {Sch{\" o}nberner}, {Steffen} \&
  {Calonaci}}{{Perinotto} et~al.}{2004}]{Perinotto2004a}
{Perinotto} M.,  {Sch{\" o}nberner} D.,  {Steffen} M.,    {Calonaci} C.,  2004,
  A\&A, 414, 993

\bibitem[\protect\citeauthoryear{{Pilyugin}}{{Pilyugin}}{2001}]{Pilyugin2001}
{Pilyugin} L.~S.,  2001, A\&A, 374, 412

\bibitem[\protect\citeauthoryear{{Pottasch}}{{Pottasch}}{1996}]{Pottasch1996}
{Pottasch} S.~R.,  1996, Ap\&SS, 238, 17

\bibitem[\protect\citeauthoryear{{Putman}, {Thom}, {Gibson} \&
  {Staveley-Smith}}{{Putman} et~al.}{2004}]{Putman2004}
{Putman} M.~E.,  {Thom} C.,  {Gibson} B.~K.,    {Staveley-Smith} L.,  2004,
  ApJ, 603, L77

\bibitem[\protect\citeauthoryear{{Rauch}}{{Rauch}}{2003}]{Rauch2003}
{Rauch} T.,  2003, A\&A, 403, 709

\bibitem[\protect\citeauthoryear{{Ruffle}, {Zijlstra}, {Walsh}, {Gray},
  {Gesicki}, {Minniti} \& {Comeron}}{{Ruffle} et~al.}{2004}]{Ruffle2004}
{Ruffle} P.~M.~E.,  {Zijlstra} A.~A.,  {Walsh} J.~R.,  {Gray} M.~D.,  {Gesicki}
  K.,  {Minniti} D.,    {Comeron} F.,  2004, MNRAS, 353, 796

\bibitem[\protect\citeauthoryear{{Saviane}, {Held} \& {Bertelli}}{{Saviane}
  et~al.}{2000}]{Saviane2000}
{Saviane} I.,  {Held} E.~V.,    {Bertelli} G.,  2000, A\&A, 355, 56

\bibitem[\protect\citeauthoryear{{Sch{\" o}nberner}, {Jacob}, {Steffen},
  {Perinotto}, {Corradi} \& {Acker}}{{Sch{\" o}nberner}
  et~al.}{2005}]{Schoenberner2005}
{Sch{\" o}nberner} D.,  {Jacob} R.,  {Steffen} M.,  {Perinotto} M.,  {Corradi}
  R.~L.~M.,    {Acker} A.,  2005, A\&A, 431, 963

\bibitem[\protect\citeauthoryear{{Soker}}{{Soker}}{2002}]{Soker2002}
{Soker} N.,  2002, MNRAS, 330, 481

\bibitem[\protect\citeauthoryear{{Stanghellini}, {Shaw}, {Balick}, {Mutchler},
  {Blades} \& {Villaver}}{{Stanghellini} et~al.}{2003}]{Stanghellini2003}
{Stanghellini} L.,  {Shaw} R.~A.,  {Balick} B.,  {Mutchler} M.,  {Blades}
  J.~C.,    {Villaver} E.,  2003, ApJ, 596, 997

\bibitem[\protect\citeauthoryear{{Stasi{\' n}ska}, {Gr{\" a}fener}, {Pe{\~
  n}a}, {Hamann}, {Koesterke} \& {Szczerba}}{{Stasi{\' n}ska}
  et~al.}{2004}]{Stasinska2004}
{Stasi{\' n}ska} G.,  {Gr{\" a}fener} G.,  {Pe{\~ n}a} M.,  {Hamann} W.-R.,
  {Koesterke} L.,    {Szczerba} R.,  2004, A\&A, 413, 329

\bibitem[\protect\citeauthoryear{{van Loon}}{{van Loon}}{2000}]{VanLoon2000}
{van Loon} J.~T.,  2000, A\&A, 354, 125

\bibitem[\protect\citeauthoryear{{van Loon}, {Cioni}, {Zijlstra} \&
  {Loup}}{{van Loon} et~al.}{2005}]{VanLoon2005}
{van Loon} J.~T.,  {Cioni} M.-R.~L.,  {Zijlstra} A.~A.,    {Loup} C.,  2005,
  A\&A, 438, 273

\bibitem[\protect\citeauthoryear{{Walsh}, {Dudziak}, {Minniti} \&
  {Zijlstra}}{{Walsh} et~al.}{1997}]{Walsh1997}
{Walsh} J.~R.,  {Dudziak} G.,  {Minniti} D.,    {Zijlstra} A.~A.,  1997, ApJ,
  487, 651

\bibitem[\protect\citeauthoryear{{Wareing}, {O'Brien}, {Zijlstra}, {Wright},
  {Greimel} \& {Drew}}{{Wareing} et~al.}{2006}]{Wareing2006}
{Wareing} C.,  {O'Brien} T.~O.,  {Zijlstra} A.~A.,  {Wright} N.,  {Greimel} R.,
     {Drew} J.~E.,  2006, MNRAS, 366, 387

\bibitem[\protect\citeauthoryear{{Wright}, {Corradi} \& {Perinotto}}{{Wright}
  et~al.}{2005}]{Wright2005}
{Wright} S.~A.,  {Corradi} R.~L.~M.,    {Perinotto} M.,  2005, A\&A, 436, 967

\bibitem[\protect\citeauthoryear{{Zijlstra}}{{Zijlstra}}{2001}]{Zijlstra2001}
{Zijlstra} A.~A.,  2001, Ap\&SS, 275, 79

\bibitem[\protect\citeauthoryear{{Zijlstra}}{{Zijlstra}}{2004}]{Zijlstra2004}
{Zijlstra} A.~A.,  2004, MNRAS, 348, L23

\bibitem[\protect\citeauthoryear{{Zijlstra}, {van Hoof}, {Chapman} \&
  {Loup}}{{Zijlstra} et~al.}{1994}]{Zijlstra1994}
{Zijlstra} A.~A.,  {van Hoof} P.~A.~M.,  {Chapman} J.~M.,    {Loup} C.,  1994,
  A\&A, 290, 228

\bibitem[\protect\citeauthoryear{{Zijlstra} \& {Walsh}}{{Zijlstra} \&
  {Walsh}}{1996}]{Zijlstra1996}
{Zijlstra} A.~A.,  {Walsh} J.~R.,  1996, A\&A, 312, L21

\end{thebibliography}

\end{document}